\documentclass[fleqn,usenatbib]{mnras}

\usepackage{newtxtext,newtxmath}
\usepackage[T1]{fontenc}

\DeclareRobustCommand{\VAN}[3]{#2}
\let\VANthebibliography\thebibliography
\def\thebibliography{\DeclareRobustCommand{\VAN}[3]{##3}\VANthebibliography}

\usepackage{graphicx}	% Including figure files
\usepackage{amsmath}	% Advanced maths commands
\usepackage{multirow}
\usepackage{xcolor}
\usepackage{ulem}

\newcommand{\Msun}{\,\mathrm{M}_\odot}
\newcommand{\feh}{\mathrm{[Fe/H]}}
\newcommand{\ofe}{\mathrm{[O/Fe]}}

\newcommand{\Mh}{M_{\mathrm{h}}}
\newcommand{\Mstar}{M_{\star}}

\newcommand{\kms}{\mathrm{\,km\,s^{-1}}}

\defcitealias{choksi_formation_2018}{CGL18}
\defcitealias{chen_modeling_2022}{CG22}
\defcitealias{chen_formation_2023}{CG23}

\title[Catalogue of model star clusters]{Catalogue of model star clusters in the Milky Way and M31 galaxies}

\author[Y. Chen and O. Y. Gnedin]{Yingtian Chen\thanks{E-mail: ybchen@umich.edu} \href{https://orcid.org/0000-0002-5970-2563}{\includegraphics[scale=0.3]{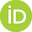}} and
Oleg Y. Gnedin \href{https://orcid.org/0000-0001-9852-9954}{\includegraphics[scale=0.3]{figures/orcid.png}}
\\
Department of Astronomy, University of Michigan, Ann Arbor, MI 48109, USA
}

\date{Accepted XXX. Received YYY; in original form ZZZ}

\pubyear{2023}

\begin{document}
\label{firstpage}
\pagerange{\pageref{firstpage}--\pageref{lastpage}}
\maketitle

\begin{abstract}
Detailed understanding of the formation and evolution of globular clusters (GCs) has been recently advanced through a combination of numerical simulations and analytical models. We employ a state-of-the-art model to create a comprehensive catalogue of simulated clusters in three Milky Way (MW) and three Andromeda (M31) analogue galaxies. Our catalogue aims to connect the chemical and kinematic properties of GCs to the assembly histories of their host galaxies. We apply the model to a selected sample of simulated galaxies that closely match the virial mass, circular velocity profile, and defining assembly events of the MW and M31. The resulting catalogue has been calibrated to successfully reproduce key characteristics of the observed GC systems, including total cluster mass, mass function, metallicity distribution, radial profile, and velocity dispersion. We find that clusters in M31 span a wider range of age and metallicity, relative to the MW, possibly due to M31's recent major merger. Such a merger also heated up the \textit{in-situ} GC population to higher orbital energy and introduced a large number of \textit{ex-situ} clusters at large radii. Understanding the impacts of galaxy mergers and accretion on the GC populations is crucial for uncovering the galaxy assembly histories.
\end{abstract}

\begin{keywords}
globular clusters: general -- Galaxy: formation --  galaxies: star clusters: general -- galaxies: individual: M31
\end{keywords}

%%%%%%%%%%%%%%%%% BODY OF PAPER %%%%%%%%%%%%%%%%%%

\section{Introduction}
\label{sec:intro}

Over the past decade, significant observational efforts have been made to uncover the origins of globular clusters (GCs) in the local universe. Spectroscopic surveys like APOGEE \citep{majewski_apache_2017} have offered valuable insights into the chemical compositions of these ancient stellar clusters. Particularly, the advent of the \textit{Gaia} mission \citep{gaia_collaboration_gaia_2016,gaia_collaboration_gaia_2018,gaia_collaboration_gaia_2023}  has revolutionized our understanding of the GC spatial and kinematic properties, and their stellar populations. Studies such as \citet{massari_origin_2019} and \citet{malhan_global_2022} took advantage of the \textit{Gaia} data to paint a comprehensive picture of the origins of Galactic GCs, shedding light on key questions related to the formation and evolutionary history of GCs, including where GCs formed and how they were brought to their current locations in the galaxy.

At the same time, numerous observational and theoretical studies have improved our understanding of the formation history of the Milky Way (MW) and Andromeda (M31) galaxies. For example, \citet{deason_progenitors_2015} investigated the ratio of blue straggler stars to blue horizontal branch stars in the MW halo and suggested the accretion of massive satellite galaxies as progenitors of the stellar halo. Inspired by this, subsequent chemical and kinematic studies focused on disk and halo stars \citep{belokurov_co-formation_2018,helmi_merger_2018,deason_apocenter_2018} and confirmed the likely major merger at lookback time $\gtrsim10$~Gyr. The progenitor galaxy of this merger is referred to as the Gaia-Sausage/Enceladus (GS/E). Analysis of the kinematics of MW stars \citep{belokurov_dawn_2022} showed that after this early bursty star forming epoch, the Galaxy transitioned to a steady stage of disk formation within $1-2$~Gyr. Cosmological simulations of galaxy formation also supported the early bursty stage and the subsequent transition to a steady state \citep{yu_born_2023,semenov_formation_2023}. On the other hand, recent studies found that M31 galaxy had a distinct assembly history from the MW, characterized by a massive merger with an M32-like progenitor around 2~Gyr ago \citep{dsouza_andromeda_2018}.

Our understanding of the formation and evolution of star clusters throughout cosmic time has also been greatly advanced over the past decade. Some of the young massive clusters formed in high-redshift galaxies may survive tidal disruption, becoming GCs at the present. In this framework, theoretical modelling of GCs has become possible by embedding necessary recipes of cluster formation and evolution as sub-grid prescriptions into cosmological simulations. For example, \citet{li_star_2017,li_star_2018} and \cite{li_star_2019} treated forming star clusters effectively as sink particles centered on giant molecular clouds in a suite of zoom-in hydrodynamical simulations. The E-MOSAICS project \citep{pfeffer_e-mosaics_2018,kruijssen_e-mosaics_2019} applied the MOSAICS model for cluster formation and evolution \citep{kruijssen_photometric_2008,kruijssen_evolution_2009,kruijssen_modelling_2011} to a re-run of the EAGLE simulations \citep{schaye_eagle_2015,crain_eagle_2015}. The EMP-\textit{Pathfinder} project \citep{reina-campos_introducing_2022} further updated the E-MOSAICS physics recipes with new prescriptions for cluster formation and the multi-phase interstellar medium (ISM). These works were able to match various observational properties of both young and old clusters. They also offered insights to the links between GC properties and the assembly history of their host galaxy  \citep{kruijssen_e-mosaics_2019,pfeffer_predicting_2020,reina-campos_mass_2020,trujillo-gomez_kinematics_2021}.

In addition to these direct simulations, post-processing methods have also gained traction in the field. Such methods employ analytical prescriptions for GC formation and evolution, applying them to the merger trees and particle outputs of existing cosmological simulations. Unlike full galaxy formation simulations, these models are not sensitive to the specific baryonic prescriptions used in simulations, and do not require costly re-runs. Such robustness and flexibility make them particularly advantageous when applied to large samples of galaxies. Examples of these methods are  \cite{renaud_origin_2017,creasey_globular_2019,phipps_first_2020,halbesma_globular_2020,valenzuela_globular_2021,valenzuela_galaxy_2023}, including previous versions of our model \citep{muratov_modeling_2010,li_modeling_2014,choksi_formation_2018,choksi_formation_2019,choksi_origins_2019}. 

Our model takes the halo merger tree as the input and triggers GC formation when the specific mass accretion rate exceeds a predefined threshold. Utilizing a sequence of scaling relations, the model analytically calculates the mass and metallicity of the newly-formed GCs. It also accounts for mass loss due to stellar evolution and tidal disruption. Starting in \citet[][hereafter \citetalias{chen_modeling_2022}]{chen_modeling_2022}, we have improved the model by including the spatial and kinematic information for GCs using tracer star particles from the simulations. This enabled a comprehensive comparison between model predictions and the 9-dimensional (mass $+$ age $+$ metallicity $+$ 3D positions $+$ 3D velocities) characteristics of MW GCs. A subsequent extension by \citet[][hereafter \citetalias{chen_formation_2023}]{chen_formation_2023} incorporated dark matter (DM) particles to represent GCs in collisionless simulations. Our refined model successfully replicated key statistics of observed GC systems, including the distributions of GC mass, metallicity, distance from the galaxy center, velocity dispersion and anisotropy. By applying this model across a wide range of galaxy mass, we reproduced global scaling relations such as the effective radius--galaxy mass and the nearly linear GC mass--galaxy mass correlations.

In this work we use the model results to produce a catalogue of model GCs with the properties that match the observations of the GC systems in MW and M31. The primary use of this catalogue is to help the analysis of observations to reveal the relationship between GC features and the assembly histories of the MW and M31. To achieve this, we start by selecting a set of simulated galaxies that meet all major observational constraints, including the virial mass, circular velocity profile, and the defining assembly events, such as the GS/E-like merger for the MW and the M32-progenitor-like merger for M31. Next, we calibrate our model parameters by comparing observed GC attributes with those in our model systems. These attributes include total GC number, mass function, metallicity distribution, radial profile, and velocity dispersion. The resulting catalogue can enhance our understanding of how galaxy assembly events influence the current distribution of GCs in the property space, such as the age--metallicity plane and the integral of motion (IOM) space. The catalogue enables us to assess the accuracy of various classification algorithms in these spaces to identify the original progenitors for \textit{ex-situ} GCs and, ultimately, reconstruct the host galaxy's assembly history.

The rest of the paper is organized as follows. We outline the background simulations and detail all updates made to the \citetalias{chen_modeling_2022} and \citetalias{chen_formation_2023} models in Sec.~\ref{sec:model_setup}. Next, we define the criteria used to select the best-matching MW and M31 analogues within our simulations in Sec.~\ref{sec:selection}. In Sec.~\ref{sec:selected_analogues}, we present the GC systems in the selected analogues and compare them with the observational data. We also explore the influence of the last major merger in M31 on the spatial distribution of GCs. The discussion in Sec.~\ref{sec:discussion} focuses on the potential application of the model to uncover galaxy assembly histories and addresses known caveats. We conclude and summarize the study in Sec.~\ref{sec:summary}. In addition, we present a new functional form for the time-dependent galactic stellar mass--metallicity relation (MZR), used to determine the metallicity of model GCs, in Appendix~\ref{sec:mzr}.

\section{Model setup}
\label{sec:model_setup}

To create model catalogues of GC systems in the MW and M31, we apply our GC formation model to 6 carefully selected galaxies from the cosmological Illustris TNG50-1 simulation \citep[][hereafter TNG50]{nelson_first_2019,pillepich_first_2019,nelson_illustristng_2021} and from a suite of collisionless simulations of a Local Group (LG) environment \citepalias{chen_formation_2023}. These simulated galaxies have assembly histories similar to the MW or M31 and reproduce most of the observable properties of their GC systems. In this section, we provide an overview of the simulations and the model setup. We describe the criteria used to select the best MW/M31 analogues in the next section.

\subsection{Background cosmological simulation}

We apply our model on two suites of simulations. The first suite is TNG50, performed with the moving mesh hydrodynamic code \textsc{arepo} \citep{springel_e_2010}. TNG50 employs the IllustrisTNG model of galaxy formation \citep{pillepich_simulating_2018} in a $(51.7\ {\rm Mpc})^3$ comoving box, adopting a flat $\Lambda$CDM cosmology with $\Omega_{\rm b}=0.0486$, $\Omega_{\rm m}=0.3089$, $\Omega_\Lambda=0.6911$, $h=0.6774$, $\sigma_8=0.8159$, and $n_{\rm s}=0.9667$ \citep{planck_collaboration_planck_2016}. TNG50 is initiated with $2160^3$ DM particles and the same number of gas particles, yielding an average DM particle mass $\sim4.5\times10^5\Msun$ and an average gas cell mass $\sim8.5\times10^4\Msun$. The gas cells in star-forming regions have typical sizes $\sim100\ {\rm pc}$ \citep{pillepich_first_2019}.

TNG50 applies the Friends-of-Friends algorithm and the \textsc{subfind} code \citep{springel_populating_2001} to identify halos and subhalos. We use the term `galaxy' to refer to subhalos throughout the paper. Additionally, TNG50 provides galaxy merger trees using the \textsc{sublink} code \citep{rodriguez-gomez_merger_2015}.

The second suite contains two collisionless zoom-in simulations, which we ran in \citetalias{chen_formation_2023}. We performed the simulations with the adaptive refinement tree \citep[ART,][]{kravtsov_adaptive_1997} code on the modified initial conditions (ICs) from the ELVIS suite \citep{garrison-kimmel_elvis_2014}. The ICs are \texttt{Thelma \& Louise} and \texttt{Romeo \& Juliet}, each producing a galaxy environment similar to the LG. We refer to this simulation suite as the `LG simulations' hereafter.

We run the \textsc{rockstar} \citep{behroozi_rockstar_2013} halo finder and the \textsc{consistent tree} code to construct halo catalogues and merger trees for the LG simulations, respectively. The two LG simulations each produces two main galaxies. We find the \texttt{Louise}, \texttt{Romeo}, and \texttt{Juliet} galaxies have a quiescent `MW-like' mass assembly history after $z\sim5$ with no major merger with a mass ratio less than 4:1. This feature is similar to the formation history of the MW \citep{hammer_milky_2007}. On the other hand, \texttt{Thelma} has more major mergers at later times.

\subsection{Modeling cluster formation and evolution}
\label{sec:cluster_model}

Our GC formation and evolution model is based on \citetalias{chen_modeling_2022}, where we modified the previous versions of the model \citep{muratov_modeling_2010,li_modeling_2014,choksi_formation_2018} to include positional and kinematic information by tagging simulation particles as `GC tracer particles'. In this section, we briefly describe the setup of the \citetalias{chen_modeling_2022} model and further modifications we make in this work.

Our model consists of four steps: 1) cluster formation, 2) cluster sampling, 3) particle assignment, and 4) cluster evolution. The formation of GCs is triggered by rapid mass growth of the host galaxy (e.g., major mergers or intense mass accretion), quantified by the specific mass accretion rate $\dot{M}_{\rm h}/\Mh$ exceeding a threshold value $p_3$, which is an adjustable model parameter. We then apply the empirical stellar mass--halo mass (SMHM) relation \citep{behroozi_average_2013} to compute the stellar mass $\Mstar$ from the halo merger history; the gas mass--stellar mass relation \citep{lilly_gas_2013,genzel_combined_2015,tacconi_phibss_2018,wang_3_2022} to calculate gas mass $M_{\rm gas}$ from stellar mass; and finally the linear gas mass--cluster mass relation \citep{kravtsov_formation_2005}, $M_{\rm tot}=1.8\times10^{-4}p_2M_{\rm gas}$, to compute the total GC mass $M_{\rm tot}$, where $p_2$ is another model parameter quantifying the cluster formation rate. We also use the time-dependent MZR to assign the host galaxy metallicity to its population of GCs forming at a given epoch. We make some modifications to the scaling relations employed in \citetalias{chen_modeling_2022} to match the updates in the theoretical and observational results and to improve the modeling of scatter evolution. We provide details of these modifications in Sec.~\ref{sec:scaling_relations}.

Next, we sample the masses of individual clusters from the \citet{schechter_analytic_1976} initial cluster mass function (ICMF) derived from observations of young star clusters. We extend the range of cluster masses down to $10^4\Msun$ instead of $10^5\Msun$ in \citetalias{chen_modeling_2022}. This allows us to capture some surviving low-mass GCs in the outskirts of the galaxies. We do not model clusters with initial mass below $10^4\Msun$ for two primary reasons. First, the lowest-mass galaxies we consider have halo masses around $10^8\Msun$ (resolved by $\sim200$ particles in TNG50), and typically produce clusters with masses close to $10^4\Msun$, see Fig.~5 in \citetalias{chen_formation_2023}. Second, our tidal disruption model is effective in disrupting clusters of this mass within a few Gyr. According to calculations in \citetalias{chen_formation_2023}, a $10^4\Msun$ cluster located at a galactocentric radius $=3$~kpc would likely survive less than 1 Gyr. Therefore, extending the mass function below this limit is meaningless.

After obtaining the list of newly formed clusters, we assign them to certain types of simulation particles depending on the simulation. For the hydrodynamic simulations like TNG50, we first select young (age $<10\ {\rm Myr}$) stellar particles within an initial radius of $3$~kpc from the galactic center. We have tested different initial radii and found that a larger radius leads to a final spatial distribution more extended than observations. However, reducing the initial radius does not significantly impact the final spatial distribution of the clusters. When there are not enough newly formed stellar particles, we also use older stellar particles formed between the adjacent snapshots. In the rare cases ($\sim$10\%) when there are still not enough stellar particles, we use DM particles near the galactic center as GC tracer particles. We only use the positions and velocities of these particles to passively track individual GCs but calculate all other properties analytically. This minimizes the model dependence on the baryon physics employed in the hydrodynamic simulation. 

On the other hand, for the collisionless LG simulations we follow \citetalias{chen_formation_2023} to select collisionless particles in local density peaks near the galaxy center, corresponding to surviving dense cores of satellite galaxies or other galactic structure with deep potential wells where massive clusters are more likely to form than elsewhere. We identify peaks within the scale radius of the best-fit Navarro–Frenk–White (NFW) halo profile. We also require the peak to be denser than any of the 16 closest grid cells and 30 times the mean density enclosed within the scale radius.

Finally, we compute the GC mass loss due to the stellar evolution and dynamical disruption based on the local tidal field along their orbits. Different from \citetalias{chen_modeling_2022}, we update the prescription for tidal disruption following \citet{gieles_mass-loss_2023}, where the disruption rate is a multivariate power-law function of the cluster initial mass and current mass. In Sec.~\ref{sec:cluster_evolution}, we describe the modifications made to the cluster evolution step in greater detail.

To choose the best model parameters, we calibrate the model to match the metallicity distribution, mass function, total GC mass--halo mass relation, and the spatial distribution. We describe the calibration in Sec.~\ref{sec:calibration}.

\subsubsection{Modifications to the scaling relations}
\label{sec:scaling_relations}

Some of the scaling relations in the \citetalias{chen_modeling_2022} model are modified. First, we modify the gas mass--stellar mass relation by updating the upper bound of the gas mass constrained by the extragalactic ultraviolet background after reionization. Following the more recent \citetalias{chen_formation_2023} model, we require the sum of the gas fraction $f_{\rm gas}=M_{\rm gas}/M_{\rm h}$ and the stellar fraction $f_*=\Mstar/M_{\rm h}$ to be less than the total accreted baryon fraction $f_{\rm in}$ parametrized by \citet{kravtsov_span_2022}:
\begin{equation}
    f_{\rm in} = f_{\rm b} \, s(M_{\rm ch}(z)/M_{\rm h}, 2),
\end{equation}
where $f_{\rm b}=\Omega_{\rm b}/\Omega_{\rm m}$ is the universal baryon fraction, $s(x,y) = [1+(2^{y/3}-1)x^y]^{-3/y}$ is a soft step-function, and $M_{\rm ch}$ is the characteristic mass scale at which $f_{\rm in}=0.5f_{\rm b}$:
\begin{equation}
    M_{\rm ch}(z) = 1.69\times 10^{10}\Msun\frac{\exp({-0.63z})}{1+\exp[(z/\beta)^\gamma]},
\end{equation}
where
\begin{equation}
    \beta = z_{\rm rei}\left[\ln\left(1.82\times 10^3\exp(-0.63z_{\rm rei})-1\right)\right]^{-1/\gamma}.
\end{equation}
We adopt the reionization epoch at $z_{\rm rei} = 6$ and $\gamma=15$ as in \citet{kravtsov_span_2022}. If $f_{\rm gas}+f_* > f_{\rm in}$, we enforce $f_{\rm gas} = f_{\rm in}-f_*$ by setting $M_{\rm gas} = (f_{\rm in}-f_*)\Mh$. 

Next, although we still use the \citet{behroozi_average_2013} SMHM relation, we change the way we model and evolve the scatter. In our previous models since \citet{choksi_formation_2018}, the scatter was modeled as the cumulative scatter of short-term star formation periods. Such a technique guarantees the stellar mass $\Mstar$ to be a monotonic function of time but may underestimate the resultant scatter at present-day. Also, this method computes the final $\Mstar$ as the summation of a series of log-normal distributions at each epoch, which may not result in a final log-normal distribution to match the present-day observations. Instead, in this work we model the evolution of scatter by a Gaussian process,
\begin{equation}
    \log\Mstar(\Mh,z)\sim{\cal GP}\left\{\log{\rm SMHM}(\Mh,z),\ K(t_1,t_2)\right\}.
    \label{eq:ms_scatter}
\end{equation}
For simplicity, we drop the `10' subscript in the base-10 logarithm for this expression and hereafter. A Gaussian process is a probabilistic model that represents a function as a probability distribution over all possible functions that are consistent with a given relation. It is fully specified by its mean function (in our case, $\log{\rm SMHM}(\Mh,z)$) and covariance function, or kernel, $K(t_1,t_2)$. The mean function provides the expected value of the function at any epoch, and the covariance function determines how the values at different epochs are related to each other. We choose a squared exponential kernel
\begin{equation}
    K(t_1,t_2) = \xi^2(z) \; \exp\left[-\frac{(t_1-t_2)^2}{2\tau^2}\right]
\end{equation}
with the scatter of the \citet{behroozi_average_2013} relation: $\xi(z)=0.218 + 0.023\, z/(1 + z)$. The parameter $\tau$ characterizes the autocorrelation timescale. The limit of $\tau\rightarrow 0$ represents a pure Gaussian noise. We set $\tau=2$ Gyr to reflect a typical gas depletion timescale in galaxies. However, we have verified that any value in the range $\tau=1-4$ Gyr does not noticeably affect any property we analyze. The Gaussian process technique solves all the caveats of the previous methods. In addition, the non-zero autocorrelation preserves memory of $\Mstar$ at previous epochs, leading to a smoother evolution of $\Mstar$. This provides a  monotonic function $\Mstar(z)$ during the peak of GC formation at $z=1-6$. At $z\lesssim1$ when the galactic star formation rate (SFR) drops and the growth of $\Mh$ slows, the calculated value of $\Mstar(z)$ may also decrease. In this case, we set $\Mstar$ to equal the previous value as we do not expect the stellar mass to decrease in reality (aside from the mass loss due to stellar evolution, which is already incorporated in the SMHM relation).

Lastly, we update the galactic MZR to the following expression:
\begin{equation}
    \feh(\Mstar,z) = 0.3\log\frac{\Mstar}{10^9\Msun} - 1.0\log(1+z) - 0.5.
    \label{eq:metalicity_stellar_mass}
\end{equation}
This relation is updated from the previous version \citep{choksi_formation_2018} by re-calibrating with the new observational data spanning a broad range of mass ($\Mstar=10^7-10^{11}\Msun$) and redshift ($z=0.7-12$). The low-$\Mstar$ and high-$z$ data are mainly from the recent programs obtained with the James Webb Space Telescope (JWST), see Appendix~\ref{sec:mzr} for a detailed description. 

Similarly to SMHM, we model the scatter of $\feh$ and its evolution via the Gaussian process,
\begin{equation}
    \feh_{\rm g}(\Mstar,z)\sim{\cal GP}\left\{\feh(\Mstar,z),\ K_{\rm g}(t_1,t_2)\right\}
    \label{eq:feh_g}
\end{equation}
where $K_{\rm g}$ is also a squared exponential function,
\begin{equation}
    K_{\rm g}(t_1,t_2) = \sigma_{\rm g}^2 \; \exp\left[-\frac{(t_1-t_2)^2}{2\tau^2}\right].
\end{equation}
The scatter of MZR is characterized by the parameter $\sigma_{\rm g}$, which we set $\sigma_{\rm g}=0.3$~dex to be consistent with the observed scatter (Appendix~\ref{sec:mzr}). Note that $\feh_{\rm g}$ refers to the mean metallicity of the galaxy. Clusters formed within this galaxy do not necessarily inherit exactly the same metallicity because of spatial variation and gradients within the ISM. We add an additional Gaussian noise to Eq.~(\ref{eq:feh_g}) to account for the internal metallicity dispersion:
\begin{equation}
    \feh_{\rm c}(\Mstar,z)\sim{\cal N}\left\{\feh_{\rm g}(\Mstar,z),\ \sigma_{\rm c}^2\right\}
    \label{eq:feh_c}
\end{equation}
where $\feh_{\rm c}$ stands for the metallicity of individual clusters formed within a galaxy of stellar mass $\Mstar$ at redshift $z$. The internal scatter is quantified by the parameter $\sigma_{\rm c}$. While $\sigma_{\rm g}$ can be measured directly from galaxy surveys, we must calibrate $\sigma_{\rm c}$ to match the observations of nearby GC systems. We achieve this by running the model with different $\sigma_{\rm c}$ to search for the model realizations that reproduce the total metallicity dispersion observed in the Virgo Cluster Survey \citep{peng_acs_2006}. We find that $\sigma_{\rm c}=0.2-0.35$~dex can match observations within one standard deviation. We adopt the smaller $\sigma_{\rm c}=0.2$~dex in this work.

\subsubsection{Modifications to cluster evolution}
\label{sec:cluster_evolution}

In the cluster evolution step, we follow the trajectories of GC particles taking into account two main processes of mass evolution: stellar mass loss and tidal disruption. Since most mass loss due to stellar evolution happens in the first tens of Myr, which is much shorter than the lifetime of a typical GC ($\sim$10~Gyr), we treat stellar evolution as an instantaneous mass loss at formation:
\begin{equation}
    M_{\rm i} = \mu_{\rm sev} \; M_0,
\end{equation}
where $M_0$ is the cluster mass at formation and $M_{\rm i}$ is the cluster mass after stellar evolution. The remaining mass fraction $\mu_{\rm sev}$ is a function of the stellar initial mass function (IMF) and metallicity. Here we assume that the IMF is constant for all clusters. Metallicity affects the exact duration of the stellar evolution but not the final remaining fraction (only by $\sim1\%$ for $-4<\feh<1$). Therefore, $\mu_{\rm sev}$ is close to a constant. We follow \citet{gieles_mass-loss_2023} to set $\mu_{\rm sev}=0.55$.

After accounting for stellar evolution, we can consider $M_{\rm i}$ as the `initial' mass before tidal disruption. Following \citetalias{chen_formation_2023} we express the tidal disruption rate of a cluster with mass $M$ as
\begin{equation}
    \frac{dM(t)}{dt} = -20\, \frac{\Msun}{\rm Myr} \left[\frac{M_{\rm i}}{2\times10^5\Msun}\right]^{1-x} \left[\frac{M(t)}{M_{\rm i}}\right]^{1-y} \left[\frac{\Omega_{\rm tid}(t)}{150\ {\rm Gyr^{-1}}}\right]
    \label{eq:mass_loss_rate}
\end{equation}
with the parameters $x=2/3$ and $y=4/3$ that match the low-density $N$-body models of \cite{gieles_mass-loss_2023}. This is different from \citetalias{chen_modeling_2022} where we used $x=y=2/3$ (which is appropriate only for very concentrated clusters).

The disruption rate also depends directly on the local tidal field strength \citep{gieles_lifetimes_2008}. We parameterize the tidal field strength by the angular frequency $\Omega_{\rm tid}$ via the effective eigenvalue $\lambda_{\rm 1,e}$ that takes into account the centrifugal, Euler, and Coriolis forces \citep{renaud_evolution_2011}:
\begin{equation}
    \Omega_{\rm tid}^2\simeq\lambda_{\rm 1,e}\simeq\lambda_1-\lambda_3
    \label{eq:omega_tid}
\end{equation}
where $\lambda_1$, $\lambda_2$, and $\lambda_3$ are the eigenvalues of the tidal tensor in descending order. This expression describes the mass loss rate more accurately than the expression in \citetalias{chen_modeling_2022}. 

We derive the tidal tensor numerically following the same method as in \citetalias{chen_modeling_2022}: we compute the second order derivatives of the gravitational potential on a $3\times 3\times 3$ cubic grid centered on the GC tracer particle, with side length $=300\ {\rm pc}$. Although the side length is still too large compared to a typical tidal radius of GCs, we cannot adopt a lower value because of limited spatial resolution of simulations. To distinguish between the true eigenvalues $\lambda$ and those we derive numerically, we use the notation $\hat{\lambda}$ for the latter. To correct for the systematic deviation of the derived value from the true $\Omega_{\rm tid}$, we use the third adjustable model parameter $\kappa$ as a correction:
\begin{equation}
    \Omega_{\rm tid}^2=\kappa \, (\hat{\lambda}_1-\hat{\lambda}_3).
\end{equation}

\section{Galaxy selection}
\label{sec:selection}

\begin{figure*}
    \includegraphics[width=0.8\linewidth]{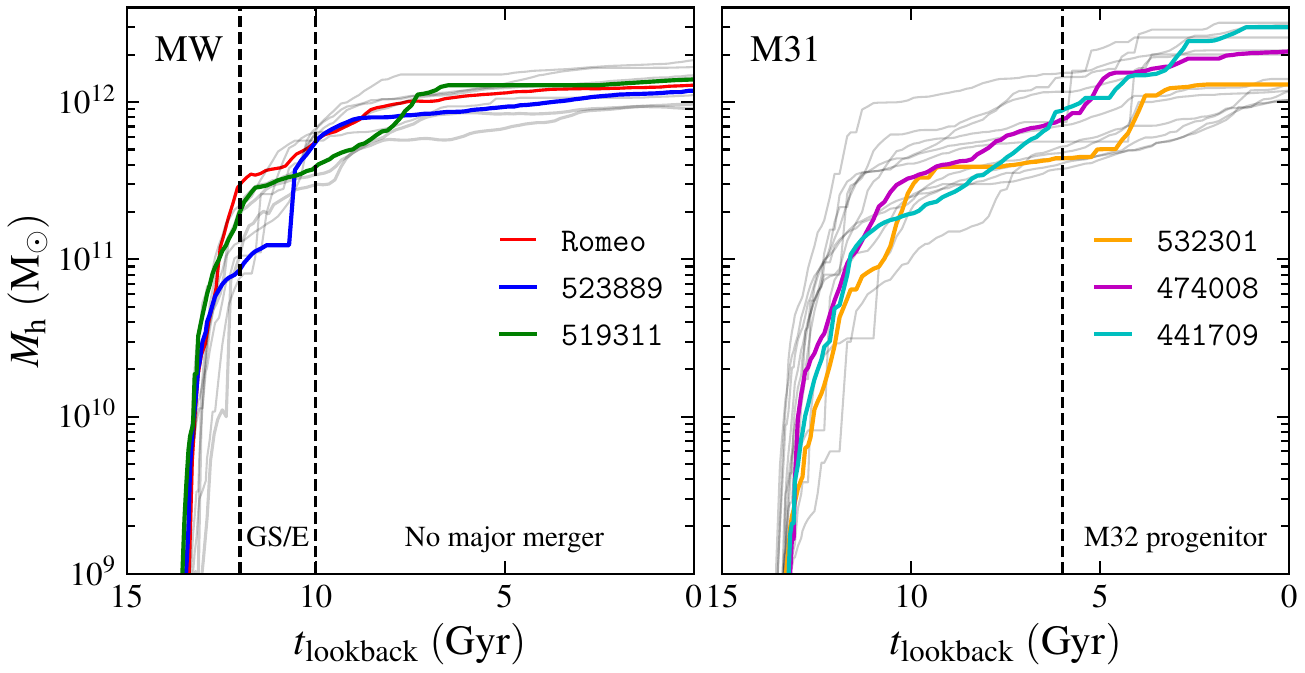}
    \vspace{-1mm}
    \caption{Mass growth histories of the main progenitor branch for the samples of 10 MW analogues (\textit{left}) and 14 M31 analogues (\textit{right}). The dashed vertical lines label the epochs for key assembly events required by our selection criteria, including the GS/E-like merger ($10-12$~Gyr) and the subsequent quiescent stage ($<10$~Gyr) for MW, and the M32-progenitor-like merger ($<6$~Gyr) for M31. We show each galaxy as a thin gray curve and highlight the three best-matching galaxies in each set with thick colored curves. We keep the same color scheme in all plots hereafter when referring to these galaxies.}
   \label{fig:mh_log_vs_tlookback}
\end{figure*}

To find GC systems in TNG50 and the LG simulations that can match the observational properties of the GC systems in the MW and M31, we first select two samples of galaxies that have properties similar to the MW and M31, respectively. Next, we apply our model to these galaxies to obtain the model GC systems at present-day. Finally, by comparing the model GC systems to the MW and M31 systems specifically, we rank the model GC systems and output the three best analogues to the MW and M31. In the following subsections, we describe the criteria to select the galaxy samples and rank model GC systems.

\subsection{Galaxy samples}

We apply the following criteria to select MW analogues in TNG50:
\begin{itemize}
    \item Galaxies with total mass\footnote{The galaxy total mass refers to the \texttt{SubhaloMass} in TNG50 \textsc{subfind} catalogue. This corresponds to the total mass of all member particles bound to this galaxy.} between $10^{11.9}$ and $10^{12.3}\Msun$.
    \item Galaxies with maximum circular velocity $V_{\rm c, max}$ between $210$ and $270\kms$.
    \item Galaxies with at least one major merger between $10-12$~Gyr ago to match the accretion of the GS/E satellite.
    \item Galaxies with no major merger in the last 10~Gyr.
    \item Galaxies formed $25-35\%$ of their present-day stellar mass \citep[calculated from halo mass via the SMHM relation of][]{behroozi_average_2013} at $t_{\rm lookback}=10$~Gyr.
\end{itemize}

Our virial mass range includes the result ($9\times10^{11}\Msun$) of the most recent modeling of the Sagittarius stream \citep{vasiliev_tango_2021}, but some deviation at the virial radius is expected because the simulated halos are not likely to have exactly the same density profile as the MW. We keep the mass variation within a factor of $\sim2$. We also add the circular velocity criterion to match the inner mass distribution, which may be more relevant for modeling star and cluster formation. The range of $V_{\rm c, max}$ is chosen to match the observed rotation curve \citep{eilers_circular_2019} with a variation $\pm 30\kms$. The circular velocity value at Sun's location at 8.5~kpc from the center is constrained to be within $200-240\kms$. The stellar mass constraint at 10~Gyr is from \citet{leitner_last_2012}.

We define a major merger as the mass ratio less than 4:1. That is, the total mass of the incoming galaxy is greater than $1/4$ of the main galaxy when the incoming galaxy reaches its maximum mass. Our two merger criteria select galaxies that assembled early. To illustrate this, we plot the mass growth histories of the main progenitor branch of these galaxies in the left panel of Fig.~\ref{fig:mh_log_vs_tlookback}. In TNG50, there are 6 galaxies that match the above criteria. Additionally, we include one more galaxy (ID \texttt{519311}\footnote{The galaxy ID refers to the \texttt{SubhaloID} in TNG50 \textsc{subfind} catalogue.}) from the sample of `early-spin-up' MW-like disk galaxies by \citet{semenov_formation_2023}. Our best MW analogue (\texttt{523889}) is also included in their sample.

We include the \texttt{Louise}, \texttt{Romeo}, and \texttt{Juliet} galaxies from the LG simulations since they have MW-like mass assembly histories. The small group environment of the LG simulations provides a more realistic background for the satellite accretion. In total, the MW sample has 10 (7 from TNG50 $+$ 3 from LG simulations) galaxy candidates with similar properties to the MW.

For the M31 sample, we follow the criteria below:
\begin{itemize}
    \item Galaxies with total mass between $10^{12}$ and $10^{12.5}\Msun$.
    \item Galaxies with at least one major merger in the recent 6~Gyr to resemble the recent merger with the M32 progenitor as suggested by \citet{dsouza_andromeda_2018}.
\end{itemize}
Since M31 contains a much richer system of GCs and its total mass is less certain than in the MW, we allow a wider range of mass. We find 14 analogues of M31 satisfying the criteria. In the right panel of Fig.~\ref{fig:mh_log_vs_tlookback} we show the mass growth histories of the main progenitor branch of these galaxies. The M31 analogues are typically assembled later and have more variable mass growth at late times compared to the MW analogues. This is directly linked to the last selection criterion. 

When selecting samples of the MW and M31 analogues, we only focus on the mass assembly history and do not take into account any baryonic properties such as the luminosity, surface brightness, or metallicity (even the stellar mass criterion for selecting MW analogues is derived from $\Mh$ using SMHM relation). This minimizes the dependence on the specific prescriptions used in the TNG50 model. However, as we show later, even the differences only in the mass assembly history can lead to a wide range of properties of the GC systems. Some of these realizations correctly match the observed properties of GC systems in the MW and M31, suggesting that GC formation is strongly related to the hierarchical assembly of galaxies.

\subsection{Model calibration}
\label{sec:calibration}

Before going into details about how we calibrate the model parameters, we introduce the observational data with which we compare the model systems. 

The mass and spatial distributions for the MW GCs are from the third version of the \citet{hilker_galactic_2019} catalogue\footnote{\url{https://people.smp.uq.edu.au/HolgerBaumgardt/globular/}}. The metallicities are from the 2010 version of the \citet{harris_catalog_1996} catalogue.

The magnitude and position data for the M31 GCs are from the Revised Bologna Catalogue\footnote{\url{http://www.bo.astro.it/M31/}} (RBC) of \citet[][version V.5, August 2012]{galleti_2mass_2004}. We compute the cluster mass from the V magnitudes using a mass-to-light ratio $M/L_{\rm V}=1.83\Msun/L_{\rm V,\odot}$ \citep{baumgardt_absolute_2020}. The metallicities are from the LAMOST spectroscopy survey of star clusters in M31 \citep{chen_lamost_2016} using spectral fitting with the models of \citet{vazdekis_evolutionary_2010}.

We limit the MW sample to clusters with $M>10^4\Msun$ to avoid the potentially incomplete low-mass clusters. For the M31 sample we require the clusters to have $M>10^{4.5}\Msun$ and locate outside the central 1~kpc (projected radius) annulus to avoid contamination near the center of M31. We also apply the same criteria to the model GC systems to consistently compare them with the observations. Since the inclination angle of the M31 galaxy is 77$^\circ$ \citep{simien_spiral_1978}, we project the model coordinates for M31 analogues using the same inclination angle with respect to the disk plane. We define the disk plane using the direction of the angular momentum of all stellar particles in the galaxy. We always refer to this inclination angle for projected radius throughout the rest of the work, unless specified otherwise.

Our model has three adjustable parameters: $p_2$, $p_3$, and $\kappa$. To find the best values for these parameters, we need to calibrate the model to match important observable features. For this purpose we use a merit function that quantifies the discrepancy between the model and observations:
\begin{equation}
    {\cal M}\equiv \frac{1}{N}\sum_{i=1}^N {\cal G}_i
\end{equation}
where ${\cal G}_i$ is a gauge function of the $i$-th simulated galaxy. We calculate the merit function separately for the MW and M31 samples. We take into account 5 features of the GC system: total GC mass, the metallicity distribution, mass function, radial distribution, and velocity dispersion. They correspond to the 5 independent terms:
\begin{equation}
    {\cal G} \equiv -\chi_M^2 - \chi_\sigma^2 + \sum_{x}\sum_{{\rm th}}\theta(p_x-p_{\rm th}).
    \label{eq:p_value}
\end{equation}

The first term is the $\chi^2$ function to evaluate the total mass of surviving clusters,
\begin{equation}
    \chi_M^2 = \frac{\left(\log M_{\rm GC}-\log M_{\rm GC,obs}\right)^2}{\sigma_M^2}
\end{equation}
where $M_{\rm GC}$ represents the total GC mass of the $i$-th galaxy, and $M_{\rm GC,obs} = 3.7\times10^7\Msun$ when compared with the MW. This number comes from the sum of individual MW cluster masses greater than $10^4\Msun$. However, since the M31 data are incomplete below $10^{4.5}\Msun$, we cannot simply sum up the mass of GCs greater than $10^4\Msun$. Instead, we fit the M31 mass function greater than $10^{4.5}\Msun$ with a log-normal function, and integrate this function down to $10^4\Msun$ to get the total GC mass. This gives a correction $\lesssim 2\%$ and yields $M_{\rm GC,obs} = 9.5\times10^7\Msun$ for M31. The denominator is the uncertainty of the number of clusters. We set $\sigma_M=0.2$~dex to approximate the uncertainty of total GC mass.

The second term is similar to the first term but evaluates the velocity dispersion of GCs,
\begin{equation}
    \chi_\sigma^2 = \frac{\left(\log \sigma_{\rm GC}-\log \sigma_{\rm GC,obs}\right)^2}{\sigma_\sigma^2}
\end{equation}
where $\sigma_{\rm GC}$ represents the 3D velocity dispersion of all surviving GCs in $i$-th galaxy, and $\sigma_{\rm GC,obs} = 200\ {\rm km\,s^{-1}}$ (calculated from the \citealp{hilker_galactic_2019} catalogue of Galactic GCs) when compared with to MW. In M31, due to background contamination near the galaxy center, most velocity measurements are limited to the outer region. For example, \citet{mackey_two_2019} studied the kinematics of clusters $\gtrsim25$~kpc and found $\sigma_{\rm GC,obs} = 134\ {\rm km\,s^{-1}}$ at $R=25$~kpc. To properly compare the model M31 analogues with the observations, we only calculate $\sigma_{\rm GC}$ of the 32 GCs closest to 25~kpc. The intrinsic scatter of the dispersion can be approximated as $\sigma_\sigma=0.3$~dex.

Inside the summation, $p_x\in\{p_M,p_Z,p_R\}$ stands for the Kolmogorov-Smirnov (KS) $p$-value for the mass function, metallicity distribution, and radial distribution. The radial distribution refers to the 3D galactocentric radius when calibrating for the MW. It refers to the projected radius for the M31. The Heaviside $\theta$ function returns 1 if $p_x$ is greater than a threshold value $p_{\rm th}$, otherwise 0. We employ $p_{\rm th}\in\{0.1,0.03,0.01\}$. In other words, $p_x>0.1$ contributes 3 points to the final ${\cal G}$, $0.03<p_x<0.1$ contributes 2 points, and $0.01<p_x<0.03$ contributes 1 point. For example, the greatest possible ${\cal G}$ is 9 if all $p_M$, $p_Z$, $p_R>0.1$, $M_{\rm GC}=M_{\rm GC,obs}$ and $\sigma_{\rm GC}=\sigma_{\rm GC,obs}$. We employ three threshold values instead of one (as in our previous model versions) to provide a finer grading hierarchy with more possible grades contributed by the three KS tests. This finer grading system avoids the discreteness that too many galaxies receive the same grade, and thus allows more nuanced calibration and selection.

For any set of model parameters, we can compute ${\cal G}$ of all MW analogues by comparing their properties to the MW. The average of ${\cal G}$ yields the merit function for the MW sample. Similarly, we can obtain the merit function for the M31 sample by comparing to M31. We then span the parameter space and select the parameters that maximize the merit functions.

It is worth noting that due to the different numerical resolution and output frequencies in TNG50 and the LG simulations, the best model parameters for the two sets of simulations are not necessarily the same. Therefore, we calibrate the model parameters separately for the two simulation sets. For TNG50, we first compute the merit functions on a grid of model parameters. The best parameters for the MW and M31 differ slightly but systematically: the merit function of the M31 maximizes at greater $p_2$ and smaller $\kappa$ compared to that of the MW. However, the merit function flattens near the maximum, allowing us to vary the parameters near the peak values without significantly affecting the model performance. We then select a single set of best parameters between the peak values of both merit functions for the MW and M31 samples: $(p_2,p_3,\kappa)=(18,0.5\ {\rm Gyr^{-1}},1.5)$.

Since the LG simulations only contain MW analogues, we simply compute the merit function for the three MW-like galaxies and select the parameters that maximize the merit function. This gives $(p_2,p_3,\kappa)=(14,0.5\ {\rm Gyr^{-1}},1.5)$. These values differ from our previous values \citepalias[][which has $p3=0.7\ {\rm Gyr^{-1}}$]{chen_formation_2023} because we have updated the model setups, especially the new prescription for cluster evolution, which effectively leads to stronger tidal disruption. To balance the decrease of clusters due to disruption, we need a smaller $p_3$ to enable more GC formation at later times.

\section{Selected MW and M31 analogues}
\label{sec:selected_analogues}

\begin{figure*}
    \includegraphics[width=0.7\linewidth]{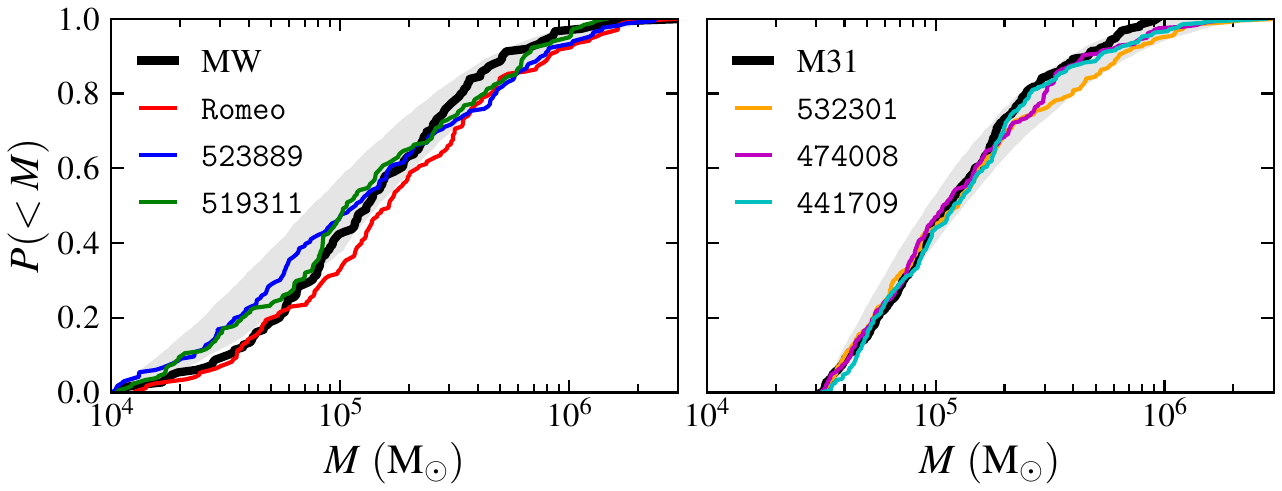}
    \includegraphics[width=0.7\linewidth]{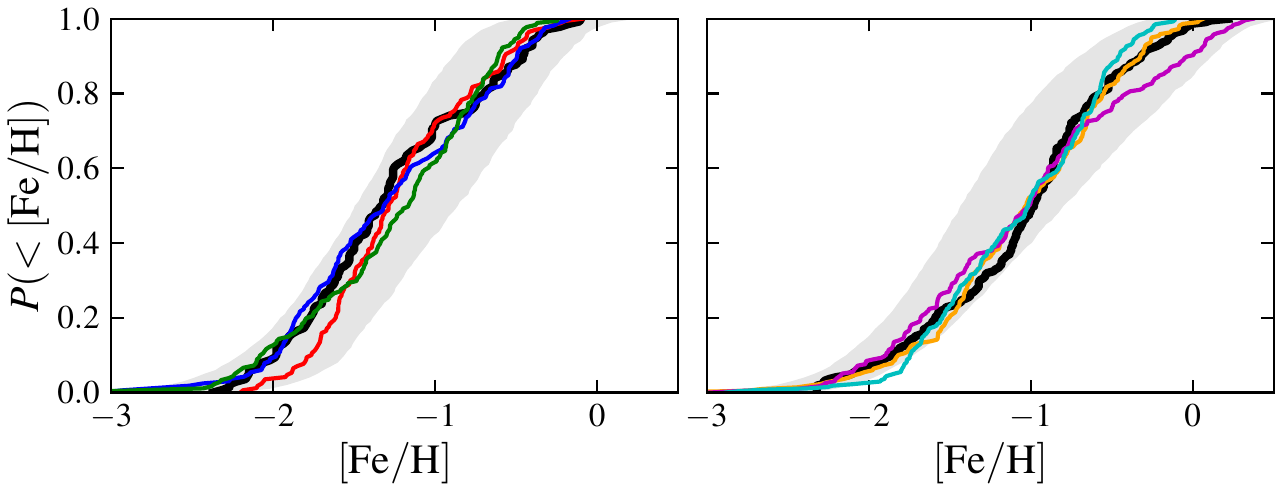}
    \includegraphics[width=0.7\linewidth]{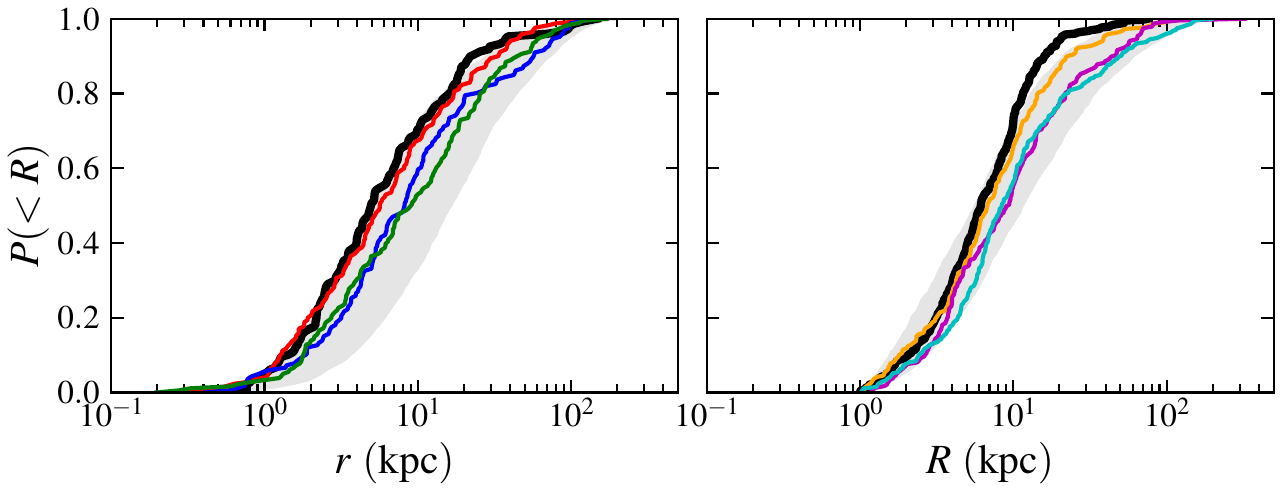}
    \vspace{-1mm}
    \caption{Cumulative distributions of the cluster mass (\textit{upper row}), metallicity (\textit{middle row}), and distance from the center (projected for M31; \textit{lower row}) for the 50 realizations of 10 MW analogues (\textit{left column}) and 14 M31 analogues (\textit{right column}), represented by the gray shaded regions as the $16-84$th percentiles. We highlight the three best-matching galaxies in each panel with the colored curves. For comparison, we overplot the observational data for the MW and M31 systems as thick black curves.}
   \label{fig:p_cum}
\end{figure*}

We run our model 50 times on each model galaxy with different random seeds to generate an ensemble of 50 realizations. Having multiple random realizations offers two major advantages. First, the model randomness, which includes the scatter in scaling relations and the stochasticity involved in sampling clusters from the ICMF and assigning clusters to simulation particles, can result in GC systems with diverse properties when the same model is rerun for the same galaxy. Analyzing the average properties of these 50 realizations reveals their systematic dependence on the model's input and the model itself.

Furthermore, having multiple random realizations enables us to search for the best MW/M31 analogues from significantly larger samples. For each MW/M31 analogue, we calculate ${\cal G}$ for the 50 realizations and identify the one with the highest ${\cal G}$ as the best representative. In this section, we present the three best-matching representatives from all sample galaxies, considering these galaxies as the best MW/M31 analogues.

\begin{table*}
 \caption{Cluster properties in the catalogue.}
 \label{tab:catalogue}
 \renewcommand\arraystretch{1.1}
 \begin{tabular}{ccl}
  \hline
  Key & Unit & Description \\
  \hline
  \texttt{galaxy\_id} & & \textsc{subfind} (TNG) or \textsc{consistent tree} (LG) ID of the current central galaxy. \\
  \texttt{t\_form} & Gyr & Formation time (lookback) of the cluster. \\
  \texttt{t\_disrupt} & Gyr & Disruption time (lookback) of the cluster; $-1$ for surviving clusters. \\
  \texttt{t\_accrete} & Gyr & Accretion time (lookback) of the cluster's host satellite galaxy; $-1$ for \textit{in-situ} clusters. \\
  \texttt{log\_m\_form} & $\rm M_\odot$ & log$_{10}$ of the cluster mass at formation. \\
  \texttt{log\_m\_gc} & $\rm M_\odot$ & log$_{10}$ of the current cluster mass; $-1$ for disrupted clusters. \\
  \texttt{x}, \texttt{y}, \texttt{z} & kpc & Galactocentric coordinates, with \texttt{z} as the stellar disk (TNG) or total (LG) angular momentum axis.$^{*}$ \\
  \texttt{vx}, \texttt{vy}, \texttt{vz} & $\rm km\,s^{-1}$ & Galactocentric velocity components.$^{**}$ \\
  \texttt{Jr}, \texttt{Jz}, \texttt{Jp} & $\rm kpc\,km\,s^{-1}$ & Orbital actions $J_r$, $J_Z$, and $J_\phi$ (Eq.~(\ref{eq:actions})); $0$ for LG. \\
  \texttt{rapo}, \texttt{rperi} & kpc & Apocenter and pericenter radii; $0$ for LG. \\
  \texttt{Ep} & $\rm km^2\,s^{-2}$ & Gravitational potential by multiple expansion (Eq.~(\ref{eq:pot1})) and spline approximation; $0$ for LG. \\
  \texttt{feh} & dex & Iron abundance $\rm [Fe/H]$. \\
  \texttt{host\_id\_form} & & \textsc{sublink} or \textsc{consistent tree} ID (main leaf) for the host galaxy at cluster formation. \\
  \texttt{host\_id\_accrete} & & \textsc{sublink} or \textsc{consistent tree} ID (main leaf) for the host satellite galaxy at accretion; $-1$ for \textit{in-situ} clusters. \\
  \texttt{log\_mh\_form} & $\rm M_\odot$ & log$_{10}$ of host galaxy total mass at cluster formation. \\
  \texttt{log\_ms\_form} & $\rm M_\odot$ & log$_{10}$ of host galaxy stellar mass at cluster formation, from \citet{behroozi_average_2013} SMHM relation.$^{***}$ \\
  \texttt{log\_mh} & $\rm M_\odot$ & log$_{10}$ of current central total mass. \\
  \texttt{log\_ms} & $\rm M_\odot$ & log$_{10}$ of current central stellar mass, from \citet{behroozi_average_2013} SMHM relation.$^{***}$ \\
  \texttt{log\_ms\_k18} & $\rm M_\odot$ & log$_{10}$ of current central stellar mass, from \citet{kravtsov_stellar_2018} SMHM relation. \\
  \hline
  \multicolumn{3}{l}{$^{*}$ The galactocentric coordinates are the simulation particle positions relative to the current central galaxy, oriented \textit{face-on}.}\\
  \multicolumn{3}{l}{$^{**}$ The galactocentric velocity components are the simulation particle velocities relative to the current central galaxy, oriented \textit{face-on}.}\\
  \multicolumn{3}{l}{$^{***}$ Scatter from Eq.~(\ref{eq:ms_scatter}) is also included in stellar mass.}
 \end{tabular}
\end{table*}

\begin{figure*}
    \includegraphics[width=\linewidth]{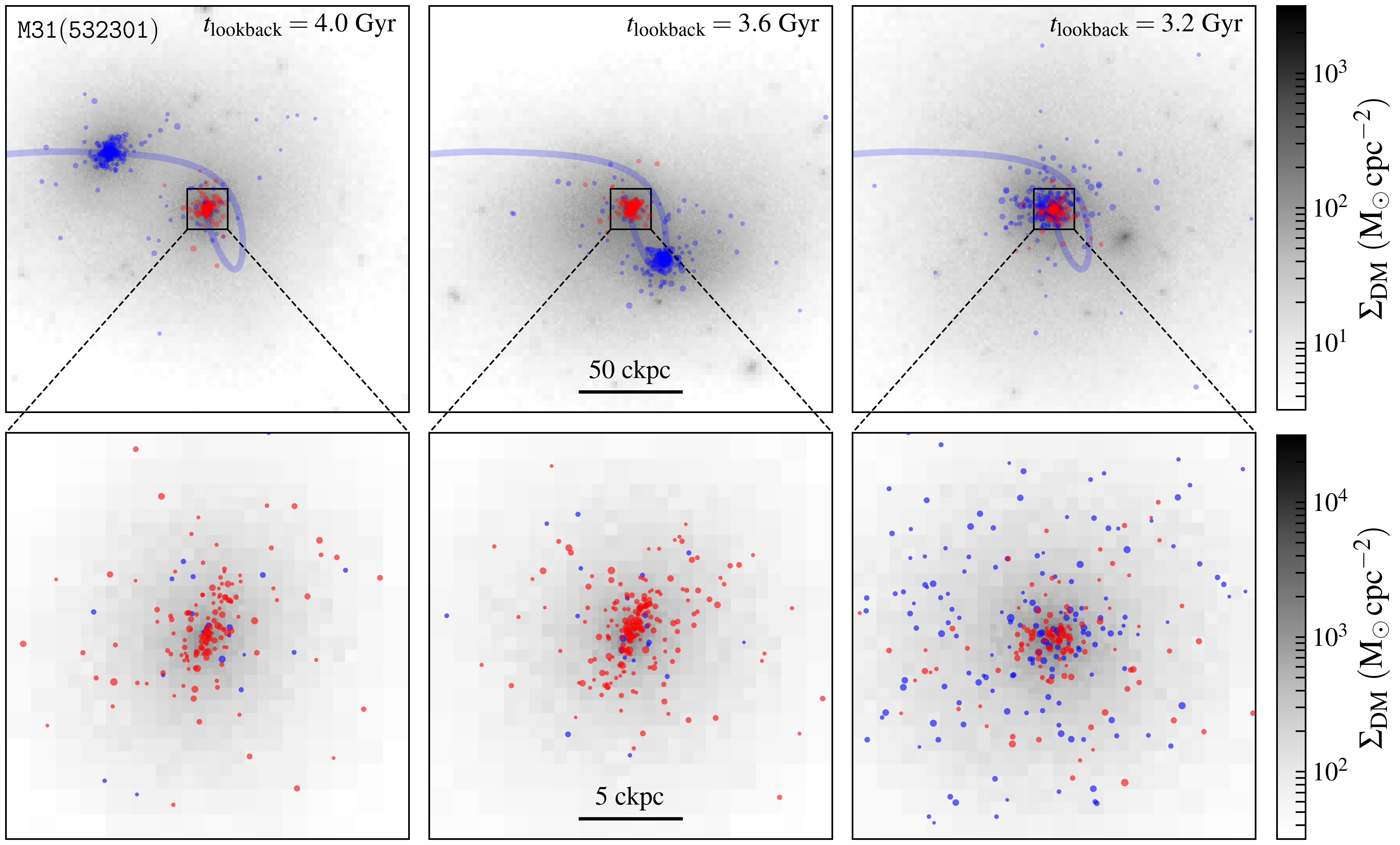}
    \vspace{-4mm}
    \caption{\textit{Face-on} projection plots of GCs during the recent major merger of the best-matching M31 analogue \texttt{532301}. From left to right, the three panels correspond to the epochs before the merger, at the satellite apocenter after the first passage, and at the end of the merger, respectively. Each panel in the \textit{top row} corresponds to a $(200\ {\rm comoving\ kpc})^3$ cube centered on the main galaxy. We plot the DM column density as background using the quadtree-based projection code \textsc{prj\_plotter} \citep{chen_prj_plotter_2023}. The \textit{in-situ} and \textit{ex-situ} clusters are shown as red and blue circles, respectively. The size of the circles increases with cluster mass. We also plot the trajectory (smoothed with cubic splines) of the incoming satellite as the blue curve. In the \textit{bottom row}, we zoom-in to the central $(20\ {\rm ckpc})^3$ region of each galaxy to show the GC distribution near the galactic center.}
   \label{fig:prj_m31}
\end{figure*}

\subsection{Properties of globular clusters in model galaxies}
\label{sec:properties}

The three best MW analogues are \texttt{Romeo} from the LG simulations and \texttt{523889} and \texttt{519311} from TNG50. The IDs of the three best M31 analogues are \texttt{532301}, \texttt{474008}, and \texttt{441709}. In Fig.~\ref{fig:mh_log_vs_tlookback}, we plot the mass growth history of these analogues as colored curves. The ${\cal G}$ function of the three M31 analogues is generally lower than that of their MW counterparts, indicating that our model is more effective in matching the MW system compared to M31. We note that the two TNG50 MW analogues and three M31 analogues are also included in the TNG50 MW/M31 sample by \citet{pillepich_milky_2023}, who selected 198 galaxies based on stellar mass, stellar morphology, and environment.

Next, we study the distributions of important GC properties. In the upper panel of Fig.~\ref{fig:p_cum} we plot the cumulative mass function for clusters in each galaxy. The GC systems of MW analogues in TNG50 have slightly more abundant low-mass clusters ($M\lesssim10^5\Msun$) compared to the observations. This is likely because the tidal disruption is underestimated for the low-mass clusters due to the limited spatial resolution of TNG50 $\sim 100$~pc. We do not find such a deviation for \texttt{Romeo} because the LG simulations have higher spatial resolution $\sim 60$~pc, and because the surviving GCs in \texttt{Romeo} are formed mainly \textit{in-situ}, which subjects them to stronger tidal disruption.

On the other hand, the mass functions of all M31 analogues are consistent with the mass function of M31 for $M>10^{4.5}\Msun$.

In the middle panel of Fig.~\ref{fig:p_cum}, we plot the cumulative cluster metallicity distribution. Our model can match the observed metallicity distributions of both MW and M31 very well. The observed MW metallicity distribution completely falls into the 1-$\sigma$ region of all MW sample galaxies. Our model also predicts that the M31 clusters have systematically broader metallicity distribution with higher abundance of metal-rich clusters compared to the MW. This is likely due to the younger GC population in M31 caused by the last major merger.

The abundance of $\feh\lesssim-1.5$ clusters in the full M31 sample is slightly higher but still within the 1-$\sigma$ region of the observations. Since $\feh\lesssim-1.5$ corresponds to GCs formed in smaller galaxies and at early epochs ($z\gtrsim3$, see Eq.~(\ref{eq:metalicity_stellar_mass})) such a mismatch may indicate the need for additional constraints on the early assembly history of M31 (our M31 selection criteria have no constraint at $t_{\rm lookback}>6$~Gyr).

Finally, we plot the cumulative distribution of distance from the galaxy center in the lower panel of Fig.~\ref{fig:p_cum}. Although the radial profile of \texttt{Romeo} can closely match the observed profile of the MW GCs with KS $p$-value $>0.1$, our model applied to TNG50 tends to produce more spatially extended GC distribution than the MW system. The half-number radii for the two best-matching MW analogues in TNG are $8-9$~kpc, which is close to twice the observational value $\sim$5~kpc. Similarly, we find that only \texttt{532301} can match the radial profile of M31 GCs. The other two best-matching M31 analogues are also spatially more extended than the M31 GC system: the half-number radius of M31 GCs is $\sim$6~kpc, whereas the two TNG50 galaxies yield $8-9$~kpc. The only model setting directly related to the spatial distribution is the initial boundary radius to distribute newly formed clusters. However, as we have verified in Sec.~\ref{sec:cluster_model}, changing this parameter does not improve the final result significantly. As we discuss later in Sec.~\ref{sec:diffuse}, this mismatch is likely because TNG50 galaxies have more abundant mergers than the galaxies in the LG environment. 

\subsection{Cluster catalogues}

Based on the above comparison, the three MW analogues and three M31 analogues are consistent with all the observable properties of their GC systems. We release the catalogues of model GCs on our model site \url{www.github.com/ognedin/gc_model_mw}. The catalogues include the following properties: galaxy ID, cluster formation/disruption/accretion time, cluster mass at formation/current, position, velocity, orbital actions, pericenter/apocenter radii, gravitational potential, metallicity, progenitor galaxy IDs, galaxy total mass at formation/current, and galaxy stellar mass at formation/current. We summarize these properties in Table~\ref{tab:catalogue} with a brief description for each entry.

Except for the orbital parameters (orbital actions, pericenter/apocenter radii, and gravitational potential) all other properties are direct outputs of the model. We describe the calculation of orbital parameters in the following section.

\subsubsection{Orbital properties}

We calculate the orbital properties of GCs using the same methods as in \citetalias{chen_modeling_2022}. We first employ the \textsc{agama} code \citep{vasiliev_agama_2019} to model the galactic potential with the multipole expansion and spline approximation methods. Since the potential of MW can be described by spheroids and disks \citep[e.g.,][]{mcmillan_mass_2017}, we model the present-day potentials of TNG50 galaxies with these two components. A largely spheroidal DM potential is modeled by the axisymmetric spherical harmonic expansion, 
\begin{equation}
    \Phi(r,\theta) = \sum_{l=0}^{l_{\rm max}}\Phi_l(r)\,Y_l^0(\theta)
    \label{eq:pot1}
\end{equation}
where $Y_l^m$ are the real-valued spherical harmonics. We only take into account the axisymmetric terms with $m=0$. The $r$-dependent coefficient $\Phi_l(r)$ is calculated on 20 grid points evenly spaced in $\log r$. The \textsc{agama} code employs a quintic spline to connect $\Phi_l(r)$ values on grid points. 

The disky baryonic (stars $+$ gas) potential is modeled as an axisymmetric form in a cylindrical coordinate system $\Phi(R,z)$. We use a 2D quintic spline to approximate the potential on a $20\times20$ grid evenly spaced in the $\log R$--$\log z$ space.

We find $l_{\rm max}=2$ is sufficient to describe the simulation-provided potential to $<2\%$ accuracy. Such an error is so small that we can ignore its influence on the subsequent calculation of the orbital parameters. 

Next, we input the present-day positions and velocities of model GCs to \textsc{agama} to calculate the orbital actions and pericenter/apocenter radii. The orbital actions of a closed orbit are defined as
\begin{equation}
    J_q \equiv \frac{1}{2\pi} \oint v_q \, dq
    \label{eq:actions}
\end{equation}
where $q\in\{r,\phi,z\}$ is the radial, azimuthal, and vertical coordinates. $J_r$ and $J_z$ characterize the oscillations in radial and azimuthal directions, whereas $J_\phi$ equals the $z$ component of angular momentum $L_z$. In separable potentials\footnote{A separable potential allows solving the Hamilton-Jacobi equation with separation of variables.}, the actions are functions of the IOMs, $J_i = J_i(I_1, I_2, I_3)$, where $I_1=E$ is the total energy. The \textsc{agama} code computes actions using the St\"{a}ckel fudge method \citep[see Sec.~3.2 in][]{vasiliev_agama_2019}. This approach assumes the potential is a St\"{a}ckel form (which is not necessarily true) so that we can analytically find the second and third integrals: the second integral in this case is $L_z$, while the third integral is non-classical. \textsc{agama} speeds up the calculation by using a pre-computed interpolation table of $J_i = J_i(I_1, I_2, I_3)$. \citet{vasiliev_agama_2019} showed that this approximation is accurate to $90-99\%$ even for extremely eccentric orbits. 

We calculate the pericenter/apocenter radii by computing the closest/farthest possible distances the cluster can reach with the current energy and angular momentum. They are the two roots of the equation $E = \Phi(r,\theta=0) + {L^2}/{2r^2}$, where $\Phi(r,\theta=0)$ is the in-plane axisymmetric potential.

We do not calculate orbital properties for \texttt{Romeo} because it is from a collisionless simulation with no baryon physics. Since the baryonic matter should become dominant near the galaxy center where most \textit{in-situ} clusters reside, the orbital parameters would be severely miscalculated. Therefore, all entries for the orbital properties in the \texttt{Romeo} catalogue are set to zero.

\subsubsection{Progenitor branch ID}

\begin{figure*}
    \includegraphics[width=\linewidth]{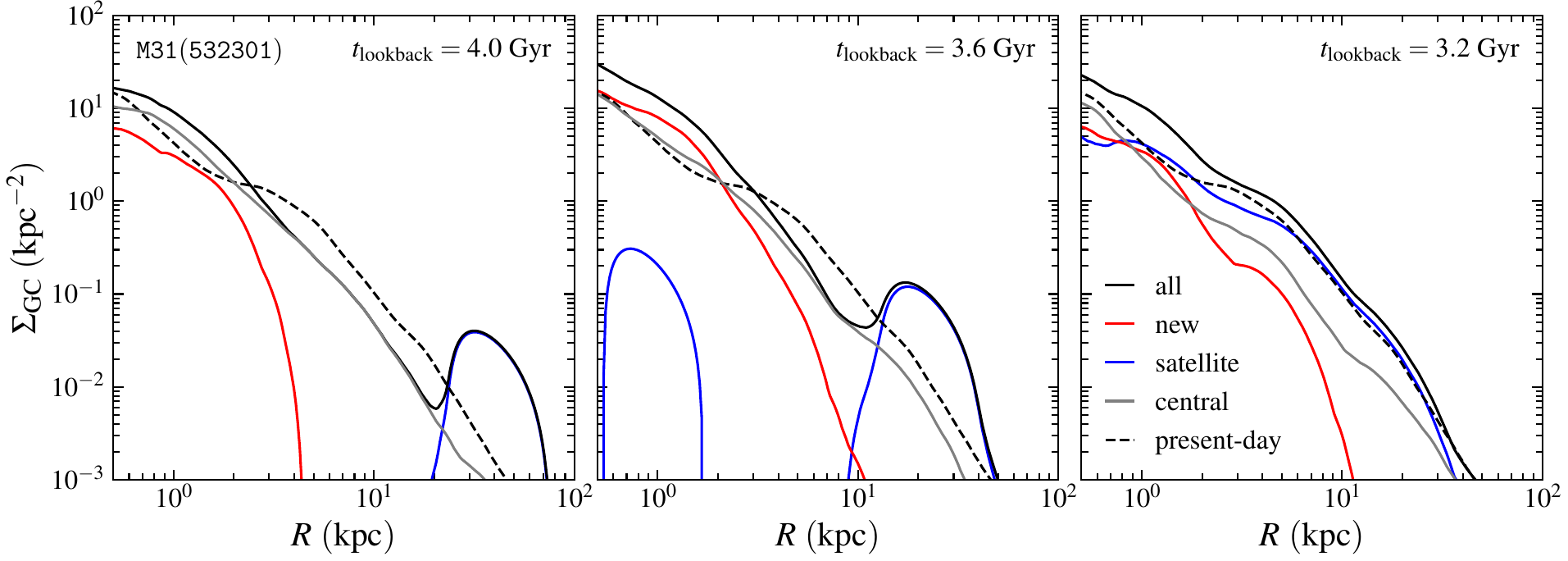}
    \vspace{-5mm}
    \caption{Evolution of the GC number density profile during the recent major merger of \texttt{532301}. From left to right, the three panels correspond to the epochs before the merger, at the satellite apocenter after the first passage, and at the end of the merger, respectively. We plot the profiles of all surviving clusters at each epoch as the solid black curve, with individual contributions of clusters brought by the satellite as blue, clusters formed in the central galaxy or accreted prior to this merger as gray, and new clusters formed during this merger as red. For comparison, we plot the present-day profile as the dashed curve.}
   \label{fig:Sigma_gc_vs_r}
\end{figure*}

In the simulation merger trees, it is common for the same galaxy at different snapshots to be assigned different IDs. It is thus more appropriate to refer to this galaxy as a `branch' rather than a `galaxy' within the terminology of merger trees. To establish a clear and unique identifier for each branch, an effective approach is to label this branch with the ID of the `main leaf' galaxy. The main leaf galaxy corresponds to the first instance of the galaxy in this branch. This labeling technique guarantees a one-to-one mapping between each branch and its main leaf ID without any ambiguity. 

We provide the \texttt{host\_id\_form} and \texttt{host\_id\_accrete} entries to identify the progenitor branches for the \textit{ex-situ} clusters. The \texttt{host\_id\_form} specifies the main leaf ID of the host galaxy at the time of cluster formation. This host galaxy may either merge directly into the central galaxy or first merge with another satellite, which is subsequently accreted onto the central galaxy. In the first case, the host galaxy at cluster formation is the same as the galaxy that ultimately delivered the cluster to the central galaxy. However, in the second scenario, the two are different. To account for this, we introduce \texttt{host\_id\_accrete} as the main leaf ID of the last satellite responsible for bringing the cluster into the central galaxy. We also provide the \texttt{t\_accrete} entry specifying its accretion time.

In a major merger event, the accreted satellite is likely to have its own merger history prior to joining the central galaxy. Since we expect all GCs from such a satellite to exhibit similar kinematics at present, regardless of their specific progenitors, the \texttt{host\_id\_accrete} entry provides a more straightforward way to query all GCs brought to the main galaxy by that satellite. Conversely, during the early stages of galaxy formation when the main galaxy was not dominant in its local environment, distinguishing all satellites becomes important. In such a scenario, the \texttt{host\_id\_form} entry is more relevant. This entry is also useful for studying the multiple populations of GCs in the same progenitor galaxy, under the assumption that GCs originating from different satellites lead to distinct populations.

\subsection{Impact of the last major merger of M31 on the cluster spatial distribution}
\label{sec:merger}

In our selection, the M31 sample is characterized by a major merger in the last 6~Gyr, whereas the MW analogues did not experience any major merger in the last 10~Gyr. The former correspond to the `late-assembled' galaxies, while the latter correspond to the `early-assembled' galaxies. In this section, we investigate the impact of the last major merger on the observable properties of the resulting GC population. Without loss of generality, we study the recent merger event of the best-matching M31 analogue \texttt{532301} as an example. All other M31 analogues lead to similar conclusions.

In Fig.~\ref{fig:prj_m31}, we plot the \textit{face-on} (not 77$^\circ$ inclination angle) projection of surviving GCs  (defined as $M>10^{4.5}\Msun$) before the merger, at the satellite apocenter after the first passage, and at the end of the merger. The merger happens at $t_{\rm lookback}=3-4$~Gyr, with a merger ratio close to unity. It can be easily noticed that the clusters brought by the recent merger have broader spatial distribution than the \textit{in-situ} clusters at the end of the merger. 

To quantify such a difference in the spatial distribution, we split the clusters into three categories: clusters formed in the satellite prior to the merger (or `satellite' clusters), clusters formed or accreted into the central galaxy prior to this merger ($t_{\rm form}>6$~Gyr, or `central' clusters), and new clusters formed during this merger ($t_{\rm form}<6$~Gyr, or `new' clusters). We plot the evolution of the \textit{face-on} radial profiles for the three categories in Fig.~\ref{fig:Sigma_gc_vs_r}. The radial profiles are obtained by kernel density estimation using an \citet{epanechnikov_non-parametric_1969} kernel: $K(x)=0.75[1-(x/h)^2]$ with $h=0.25$~dex. The peak at $R\sim1$~kpc in the middle panel is contributed by just one cluster captured by the central galaxy during the first passage. Since most satellite clusters are still bound to the satellite at this stage, we do not analyse this outlier further.

As the satellite approaches the main galaxy around $t_{\rm lookback}=3.6$~Gyr, it does not largely alter the spatial distribution of central clusters $\lesssim 10$~kpc (approximately the pericenter distance of the satellite's first passage). In contrast, it pulls the clusters in the outer region $\gtrsim 10$~kpc outwards, as the satellite is massive enough that it perturbs the potential of the central galaxy significantly in this region. The formation of about 110 new clusters increases the number density within $\lesssim 3$~kpc. However, strong tidal disruption quickly brings the numbers down by the end of the merger ($t_{\rm lookback}=3.2$~Gyr). In contrast to such dramatic early disruption, the number of new clusters only drops gradually in the next 3.2~Gyr until the present. Similarly, the early tidal disruption removes around 40 central clusters during the merger, whereas the subsequent evolution only disrupts less than 30 central clusters in the next 3.2~Gyr, leaving $\sim$90 central clusters surviving until the present.

Despite the change in the normalization, the radial profile of the central clusters does not change significantly during the merger. The merger only raises the effective radius of the central clusters from 2.5~kpc to 3.7~kpc. However, the satellite clusters have a radial profile significantly more out-spread than the centrals: the effective radius of the satellite clusters is 6.1~kpc at the end of the merger, raising the overall effective radius to 4.6~kpc. 

In the subsequent 3.2~Gyr, the radial profile does not change much, except for the inner part being lowered by tidal disruption, slightly increasing the final effective radius to 5.2~kpc.

By adding a large number of \textit{ex-situ} clusters with distinct radial distribution from the central cluster population, the recent major merger can significantly enlarge the GC effective radius by a factor of 2 in less than 1~Gyr. However, the merger does not significantly alter the spatial distribution of central clusters $\lesssim 10$~kpc. Also, the merger triggers the formation of a young \textit{in-situ} GC population within $\lesssim 3$~kpc. Combining these two factors may explain why many M31 clusters are still located around the disk without being stripped away by the merger. But as we show below in Sec.~\ref{sec:usage}, the last major merger can still heat up the \textit{in-situ} clusters to higher energy.

\section{Discussion}
\label{sec:discussion}

\begin{figure*}
    \includegraphics[width=0.7\linewidth]{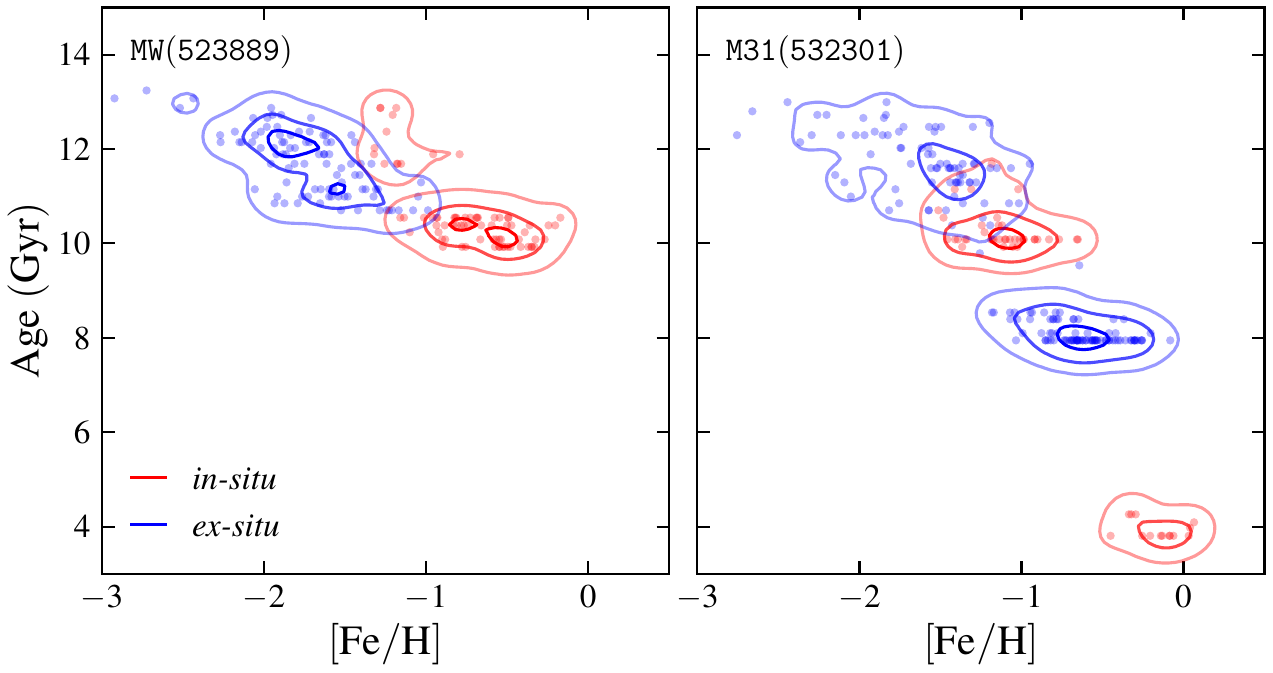}
    \vspace{-1mm}
    \caption{Cluster age-metallicity relation for a typical MW analogue (\texttt{523889}, \textit{left panel}) and a typical M31 analogue (\texttt{532301}, \textit{right panel}). Each dot represents a surviving cluster either formed \textit{in-situ} (red) or \textit{ex-situ} (blue). We calculate the contours using Gaussian KDE with bandwidth $=0.1$~dex for horizontal axis, and $0.3$~Gyr for vertical axis. The contours from dark to light enclose $10$, $50$, and $90\%$ of total number, respectively. }
   \label{fig:amr}
\end{figure*}

\subsection{Investigating galaxy assembly with the catalogue}
\label{sec:usage}

The main goal of constructing this catalogue is to investigate connections between the galaxy assembly history and the properties of GCs. For example, we show the age-metallicity relations (AMR) of surviving clusters for a typical MW analogue (\texttt{523889}) and a typical M31 analogue (\texttt{532301}) in Fig.~\ref{fig:amr}. GCs in the MW analogue are in general older than the M31 counterparts because of the early assembly history of MW. This leads to almost no GC formation after $t_{\rm lookback}=10$~Gyr. In contrast, GCs in M31 span a wider range of ages. For \texttt{532301} specifically, there are four bursts of GC formation: 1) $\sim40$ \textit{ex-situ} GCs formed at $t_{\rm lookback}\approx12$~Gyr, corresponding to active cluster formation in the early universe when galaxy mergers are frequent; 2) subsequent `bottom-up' assembly of satellites contributed a significant amount of mass to the main galaxy, leading to the formation of majority of \textit{in-situ} GCs at $\approx10$~Gyr; 3) similar growth of the largest companion galaxy at $\approx8$~Gyr, forming $\sim 130$ GCs; 4) the major merger between the main galaxy and its companion boosted \textit{in-situ} GC formation at $\approx4$~Gyr (as illustrated in Sec.~\ref{sec:merger}). Other M31 analogues also have more extended and discrete GC formation histories, but the timing and order for each GC population may differ among galaxies.

Another noticeable difference is the larger metallicity separation between \textit{in-situ} and \textit{ex-situ} clusters in the MW analogue. Because the assembly of MW is less hierarchical than that of M31, the main progenitor galaxy of MW is more dominant in the GC populations. This enlarges the gap between the metallicity of clusters formed in the main galaxy and the satellites. In contrast, most of \textit{ex-situ} clusters in the M31 analogue formed in the largest companion galaxy which had mass and metallicity comparable to the main galaxy. The metallicity of these \textit{ex-situ} clusters is thus similar to the \textit{in-situ} clusters formed before the merger.

\begin{figure*}
    \includegraphics[width=0.7\linewidth]{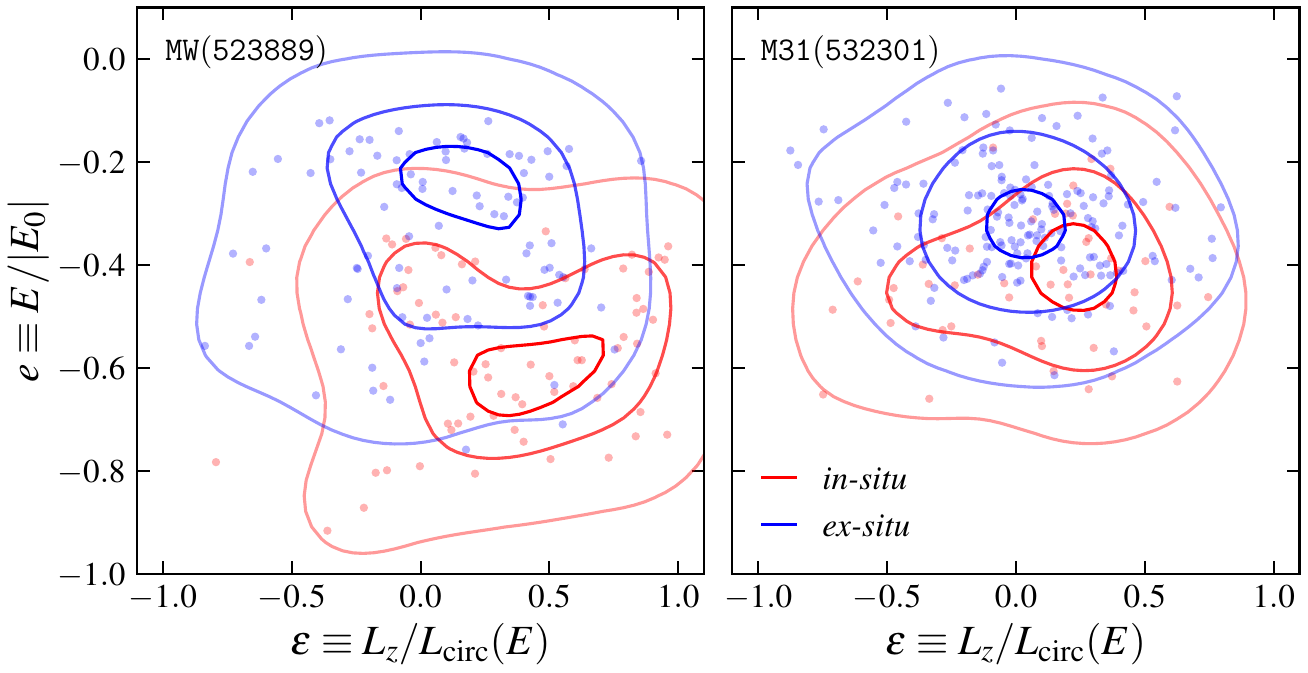}
    \vspace{-1mm}
    \caption{Normalized integral of motion space for clusters in a typical MW analogue (\texttt{523889}, \textit{left panel}) and a typical M31 analogue (\texttt{532301}, \textit{right panel}). Each dot represents a surviving cluster either formed \textit{in-situ} (red) or \textit{ex-situ} (blue). We calculate the contours using Gaussian KDE with bandwidth $=0.2$ for horizontal axis, and $0.1$ for vertical axis. The contours from dark to light enclose $10$, $50$, and $90\%$ of total number, respectively.}
   \label{fig:iom_normalize}
\end{figure*}

In addition to the AMR, galaxy assembly histories shape the orbits of GCs. For example, Fig.~\ref{fig:iom_normalize} shows the normalized IOM space for the MW and M31 analogues (\texttt{523889} and \texttt{532301}). The circularity parameter $\varepsilon$ is defined as $L_z/L_{\rm circ}(E)$, where $L_{\rm circ}(E)$ is the angular momentum of an in-plane circular orbit with the same energy $E$ as the cluster. The normalized energy $e$ is defined as $E/|E_0|$, where $E_0$ is the gravitational potential at the galaxy center. We have $\varepsilon\in[-1,1]$ and $e\in[-1,0)$. Perfectly circular and in-plane orbits have $\varepsilon=1$ (prograde) or $\varepsilon=-1$ (retrograde), while purely radial or polar orbits have $\varepsilon=0$. 

We find that the \textit{ex-situ} populations in both MW and M31 analogues have similar distribution around $(\varepsilon,e)=(0,-0.3)$. However, the \textit{in-situ} populations have quite different distributions: in the MW \textit{in-situ} GCs are located near $(\varepsilon,e)=(0.5,-0.6)$, indicating significantly lower-energy and more disky orbits. The \textit{in-situ} GCs of the M31, however, have a more similar distribution to the \textit{ex-situ} population and center around $(\varepsilon,e)=(0.2,-0.4)$. Considering the significant scatter $\Delta\varepsilon=0.3-0.4$ and $\Delta e=0.1-0.15$ in each population, the distributions of \textit{in-situ} and \textit{ex-situ} clusters in the M31 analogue largely overlap in the normalized IOM space. This is likely due to the recent major merger which heated up the \textit{in-situ} clusters to higher energy and moved some of them away from the disk.

Since GCs formed in different progenitor galaxies are located in distinct regions within the chemical and kinematic property space, numerous attempts \citep[e.g.,][]{massari_origin_2019,malhan_global_2022,belokurov_-situ_2023} have been made to identify GCs of various origins by classifying isolated clusters/groups in the AMR, IOM, and chemistry spaces. However, these studies often yield conflicting results. Our catalog can serve as a tool to assess the accuracy of the classification algorithms employed in these studies since our model provides `true' labels from the simulations (as indicated by the \texttt{host\_id\_from} and \texttt{host\_id\_accrete} entries in Table~\ref{tab:catalogue}). Furthermore, our catalog offers a unique opportunity to investigate the spread of properties among GCs originating from the same progenitor and to explore how much GCs retain their original kinematic signatures during galaxy assembly. Finding the answers to these questions is crucial for characterizing the accuracy of current classification schemes.

\subsection{Why are TNG50 GC systems more extended?}
\label{sec:diffuse}

As shown in Sec.~\ref{sec:properties}, the GC systems in TNG50 galaxies tend to have more radially extended distribution than the observations: the half-number radii of the two best-matching MW analogues in TNG50 are around 1.7 times the observational value $\sim$5~kpc. In fact, all 7 TNG50 galaxies in the MW sample have greater half-number radii, with a median value $\sim$12~kpc. We find a similar trend for the M31 sample. In contrast, all three MW-like galaxies in the LG simulations can match the radial profile of MW GCs with half-number radii $\sim$5~kpc.

These extended distributions of the TNG50 systems may be a general outcome of the large-scale environment probed by the TNG50 simulation. For example, \citet{semenov_formation_2023} showed that approximately 90\% of TNG50 galaxies developed their disk later than the MW. This difference arises from TNG50 lacking a quiescent, merger-free environment, which is necessary for the survival of an early-formed disk. While we limit our MW sample galaxies to those without major mergers with a mass ratio less than 4:1 in the last 10~Gyr, we do not impose constraints on higher merger ratios ($\sim$10:1) which can still play a significant role in structure formation.

Additional evidence indicating that MW may be less merger-dominated than typical TNG50 galaxies lies in the high \textit{in-situ} fraction of MW GCs. Observational classification studies suggest that the \textit{in-situ} fraction may range from about 40\% \citep{massari_origin_2019} to 56\% \citep{malhan_global_2022} and up to 66\% \citep{belokurov_-situ_2023}. In contrast, the TNG50 galaxies in the MW sample typically have \textit{in-situ} fractions between 20\% and 50\%, once again suggesting that even the selected MW sample galaxies may be too merger-dominated compared to the real MW.

The relatively low fraction of \textit{in-situ} clusters in TNG50 leads to two discrepancies between the model and observed data. First, the low \textit{in-situ} fraction accounts for the too extended distribution of TNG GCs, as shown in the lower left panel of Fig.~\ref{fig:p_cum}. This is because the median radius of \textit{ex-situ} clusters is approximately four times larger than that of the \textit{in-situ} clusters \citepalias[see][]{chen_modeling_2022}. Second, the low \textit{in-situ} fraction in TNG50 also explains why the AMR of TNG GCs lacks a sequence of old, metal-poor \textit{in-situ} clusters (as shown in the left panel of Fig.~\ref{fig:amr}), which is a robust feature of the MW assembly history. To align the simulations more closely with observed data, a more realistic simulation of the MW assembly is needed, particularly to match the observed AMR of the \textit{in-situ} clusters.

\section{Summary}
\label{sec:summary}

We introduce a catalogue of model star clusters in the MW and M31 analogues, drawn from the TNG50 and LG simulations. By applying criteria based on galaxy virial mass, circular velocity profile, and defining assembly events, we select a sample comprising 10 MW analogues and 14 M31 analogues. We then apply an analytical model of GC formation and evolution to the merger trees and particle outputs of these simulated galaxies. This enables us to obtain key observables such as the mass, age, metallicity, positions, and velocities of the surviving cluster population. We calibrate the model parameters to optimize agreement with the observed total cluster mass, mass function, metallicity distribution, radial profile, and velocity dispersion. From these model results we select the three best-matching GC systems each for MW and M31. One of the best-matching MW analogues is from the LG simulations; the other five are from TNG50 (see Fig.~\ref{fig:mh_log_vs_tlookback} for the mass growth histories of the six galaxies).

The GC systems in the three best-matching MW analogues successfully reproduce the observed mass function and metallicity distribution (Fig.~\ref{fig:p_cum}) after model calibration. The galaxy from the LG simulations also matches the observed radial distribution, yielding a KS $p$-value $>0.1$. However, the radial profiles of the other two GC systems from TNG50 are more extended, with effective radii approximately 1.7 times the observed value. Likewise, the three best-matching M31 analogues have mass and metallicity distributions that agree with observational data. However, only one M31 analogue reproduces the observed radial distribution, while the remaining two exhibit more extended profiles, with effective radii $\sim$1.5 times the observed value. This discrepancy may arise from the higher rate of mergers in TNG50 compared to the LG environment. Such mergers typically introduce a significant population of \textit{ex-situ} clusters, which tend to be located $\sim$4 times father away from the galaxy center than their \textit{in-situ} counterparts.

An in-depth study of the last major merger in M31 analogues also supports the above conclusion. We find that even a 1:1 merger has a limited impact on the spatial distribution of \textit{in-situ} clusters within $\lesssim10$~kpc (Fig.~\ref{fig:Sigma_gc_vs_r}). However, such a merger can notably enlarge the overall GC effective radius by a factor of 2 in a short period $\lesssim1$ Gyr, primarily by bringing in a large number of \textit{ex-situ} clusters. Moreover, the merger triggers the formation of a new population of young \textit{in-situ} clusters near the galactic center. This population, in addition to older clusters brought in by the merger, forms distinct groups in the AMR (Fig.~\ref{fig:amr}). These groups have lower age and higher metallicity compared to the central clusters formed prior to the merger, resulting in an AMR spanning a wide range in both age and metallicity. 

Furthermore, we show that the galaxy assembly history significantly influences GC kinematics. The MW analogues have a quiescent formation history over the past 10~Gyr. Such a long period preserves the original location of clusters in the IOM space (Fig.~\ref{fig:iom_normalize}). In this space, \textit{in-situ} clusters tend to have lower energy and angular momentum $L_z$ closer to $L_{\rm circ}(E)$, whereas \textit{ex-situ} clusters generally have higher energy and $L_z\approx 0$. This distinction is particularly helpful when using the GC properties to decode the merger history of their host galaxy. Conversely, the last major merger of M31 elevates the \textit{in-situ} populations to higher energy, making the \textit{in-situ} and \textit{ex-situ} populations less distinguishable in the IOM space.

We make our catalogue available in a machine-readable format. Along with all the variables directly output by the model, the catalogue also includes several orbital parameters: the gravitational potential, actions, and pericenter/apocenter radii. A comprehensive list of keys and descriptions for these variables is provided in Table~\ref{tab:catalogue}. The catalogue can be accessed at \url{www.github.com/ognedin/gc_model_mw}.

\section*{Acknowledgements}

We thank Eric Bell, Leandro Beraldo e Silva, Behzad Tahmasebzadeh, Monica Valluri, Gillen Brown, and Xi Meng for insightful discussions. This research was mainly conducted with the \textsc{python} programming language, employing the following packages: \textsc{numpy} \citep{harris_array_2020}, \textsc{matplotlib} \citep{hunter_matplotlib_2007}, \textsc{scipy} \citep{virtanen_scipy_2020}, \textsc{agama} \citep{vasiliev_agama_2019}, \textsc{yt} \citep{turk_yt_2011}, \textsc{illustris\_python} \citep{nelson_illustristng_2021} and \textsc{prj\_plotter} \citep{chen_prj_plotter_2023}. OG and YC were supported in part by the U.S. National Science Foundation through grant AST-1909063 and by National Aeronautics and Space Administration through contract NAS5-26555 for Space Telescope Science Institute program HST-AR-16614.

\section*{Data Availability}

The catalogues of model clusters in the MW and M31 galaxies are available at \url{www.github.com/ognedin/gc_model_mw}. Other data that support the findings of this study are available from the corresponding author, upon reasonable request. The TNG50 simulation data are publicly available at \url{www.tng-project.org/data}.


\bibliographystyle{mnras}
\bibliography{GC-model-references}

\begin{thebibliography}{}
\makeatletter
\relax
\def\mn@urlcharsother{\let\do\@makeother \do\$\do\&\do\#\do\^\do\_\do\%\do\~}
\def\mn@doi{\begingroup\mn@urlcharsother \@ifnextchar [ {\mn@doi@} {\mn@doi@[]}}
\def\mn@doi@[#1]#2{\def\@tempa{#1}\ifx\@tempa\@empty \href {http://dx.doi.org/#2} {doi:#2}\else \href {http://dx.doi.org/#2} {#1}\fi \endgroup}
\def\mn@eprint#1#2{\mn@eprint@#1:#2::\@nil}
\def\mn@eprint@arXiv#1{\href {http://arxiv.org/abs/#1} {{\tt arXiv:#1}}}
\def\mn@eprint@dblp#1{\href {http://dblp.uni-trier.de/rec/bibtex/#1.xml} {dblp:#1}}
\def\mn@eprint@#1:#2:#3:#4\@nil{\def\@tempa {#1}\def\@tempb {#2}\def\@tempc {#3}\ifx \@tempc \@empty \let \@tempc \@tempb \let \@tempb \@tempa \fi \ifx \@tempb \@empty \def\@tempb {arXiv}\fi \@ifundefined {mn@eprint@\@tempb}{\@tempb:\@tempc}{\expandafter \expandafter \csname mn@eprint@\@tempb\endcsname \expandafter{\@tempc}}}

\bibitem[\protect\citeauthoryear{Asplund, Amarsi  \& Grevesse}{Asplund et~al.}{2021}]{asplund_chemical_2021}
Asplund M.,  Amarsi A.~M.,   Grevesse N.,  2021, \mn@doi [A\&A] {10.1051/0004-6361/202140445}, 653, A141

\bibitem[\protect\citeauthoryear{Baumgardt, Sollima  \& Hilker}{Baumgardt et~al.}{2020}]{baumgardt_absolute_2020}
Baumgardt H.,  Sollima A.,   Hilker M.,  2020, \mn@doi [PASA] {10.1017/pasa.2020.38}, 37, e046

\bibitem[\protect\citeauthoryear{Behroozi, Wechsler  \& Wu}{Behroozi et~al.}{2013a}]{behroozi_rockstar_2013}
Behroozi P.~S.,  Wechsler R.~H.,   Wu H.-Y.,  2013a, \mn@doi [ApJ] {10.1088/0004-637X/762/2/109}, 762, 109

\bibitem[\protect\citeauthoryear{Behroozi, Wechsler  \& Conroy}{Behroozi et~al.}{2013b}]{behroozi_average_2013}
Behroozi P.~S.,  Wechsler R.~H.,   Conroy C.,  2013b, \mn@doi [ApJ] {10.1088/0004-637X/770/1/57}, 770, 57

\bibitem[\protect\citeauthoryear{Belokurov \& Kravtsov}{Belokurov \& Kravtsov}{2022}]{belokurov_dawn_2022}
Belokurov V.,  Kravtsov A.,  2022, \mn@doi [MNRAS] {10.1093/mnras/stac1267}, 514, 689

\bibitem[\protect\citeauthoryear{Belokurov \& Kravtsov}{Belokurov \& Kravtsov}{2023}]{belokurov_-situ_2023}
Belokurov V.,  Kravtsov A.,  2023, arXiv:2309.15902 [astro-ph]

\bibitem[\protect\citeauthoryear{Belokurov, Erkal, Evans, Koposov  \& Deason}{Belokurov et~al.}{2018}]{belokurov_co-formation_2018}
Belokurov V.,  Erkal D.,  Evans N.~W.,  Koposov S.~E.,   Deason A.~J.,  2018, \mn@doi [MNRAS] {10.1093/mnras/sty982}, 478, 611

\bibitem[\protect\citeauthoryear{Birkin et~al.,}{Birkin et~al.}{2023}]{birkin_jwsts_2023}
Birkin J.~E.,  et~al., 2023, arXiv:2307.10412 [astro-ph]

\bibitem[\protect\citeauthoryear{Boyett et~al.,}{Boyett et~al.}{2023}]{boyett_massive_2023}
Boyett K.,  et~al., 2023, arXiv:2303.00306 [astro-ph]

\bibitem[\protect\citeauthoryear{Chen}{Chen}{2023}]{chen_prj_plotter_2023}
Chen Y.,  2023, prj\_plotter: v0.1.2, \mn@doi{10.5281/ZENODO.8165883}

\bibitem[\protect\citeauthoryear{Chen \& Gnedin}{Chen \& Gnedin}{2022}]{chen_modeling_2022}
Chen Y.,  Gnedin O.~Y.,  2022, \mn@doi [MNRAS] {10.1093/mnras/stac1651}, 514, 4736

\bibitem[\protect\citeauthoryear{Chen \& Gnedin}{Chen \& Gnedin}{2023}]{chen_formation_2023}
Chen Y.,  Gnedin O.~Y.,  2023, \mn@doi [MNRAS] {10.1093/mnras/stad1328}, 522, 5638

\bibitem[\protect\citeauthoryear{Chen et~al.,}{Chen et~al.}{2016}]{chen_lamost_2016}
Chen B.,  et~al., 2016, \mn@doi [AJ] {10.3847/0004-6256/152/2/45}, 152, 45

\bibitem[\protect\citeauthoryear{Choksi \& Gnedin}{Choksi \& Gnedin}{2019a}]{choksi_formation_2019}
Choksi N.,  Gnedin O.~Y.,  2019a, \mn@doi [MNRAS] {10.1093/mnras/stz811}, 486, 331

\bibitem[\protect\citeauthoryear{Choksi \& Gnedin}{Choksi \& Gnedin}{2019b}]{choksi_origins_2019}
Choksi N.,  Gnedin O.~Y.,  2019b, \mn@doi [MNRAS] {10.1093/mnras/stz2097}, 488, 5409

\bibitem[\protect\citeauthoryear{Choksi, Gnedin  \& Li}{Choksi et~al.}{2018}]{choksi_formation_2018}
Choksi N.,  Gnedin O.~Y.,   Li H.,  2018, \mn@doi [MNRAS] {10.1093/mnras/sty1952}, 480, 2343

\bibitem[\protect\citeauthoryear{Crain et~al.,}{Crain et~al.}{2015}]{crain_eagle_2015}
Crain R.~A.,  et~al., 2015, \mn@doi [MNRAS] {10.1093/mnras/stv725}, 450, 1937

\bibitem[\protect\citeauthoryear{Creasey, Sales, Peng  \& Sameie}{Creasey et~al.}{2019}]{creasey_globular_2019}
Creasey P.,  Sales L.~V.,  Peng E.~W.,   Sameie O.,  2019, \mn@doi [MNRAS] {10.1093/mnras/sty2701}, 482, 219

\bibitem[\protect\citeauthoryear{Curti, Mannucci, Cresci  \& Maiolino}{Curti et~al.}{2020}]{curti_massmetallicity_2020}
Curti M.,  Mannucci F.,  Cresci G.,   Maiolino R.,  2020, \mn@doi [MNRAS] {10.1093/mnras/stz2910}, 491, 944

\bibitem[\protect\citeauthoryear{Curti et~al.,}{Curti et~al.}{2022}]{curti_chemical_2022}
Curti M.,  et~al., 2022, \mn@doi [MNRAS] {10.1093/mnras/stac2737}, 518, 425

\bibitem[\protect\citeauthoryear{Curti et~al.,}{Curti et~al.}{2023}]{curti_jades_2023}
Curti M.,  et~al., 2023, arXiv:2304.08516 [astro-ph]

\bibitem[\protect\citeauthoryear{Deason, Belokurov  \& Weisz}{Deason et~al.}{2015}]{deason_progenitors_2015}
Deason A.~J.,  Belokurov V.,   Weisz D.~R.,  2015, \mn@doi [MNRAS] {10.1093/mnrasl/slv001}, 448, L77

\bibitem[\protect\citeauthoryear{Deason, Belokurov, Koposov  \& Lancaster}{Deason et~al.}{2018}]{deason_apocenter_2018}
Deason A.~J.,  Belokurov V.,  Koposov S.~E.,   Lancaster L.,  2018, \mn@doi [ApJ] {10.3847/2041-8213/aad0ee}, 862, L1

\bibitem[\protect\citeauthoryear{D’Souza \& Bell}{D’Souza \& Bell}{2018}]{dsouza_andromeda_2018}
D’Souza R.,  Bell E.~F.,  2018, \mn@doi [Nat Astron] {10.1038/s41550-018-0533-x}, 2, 737

\bibitem[\protect\citeauthoryear{Eilers, Hogg, Rix  \& Ness}{Eilers et~al.}{2019}]{eilers_circular_2019}
Eilers A.-C.,  Hogg D.~W.,  Rix H.-W.,   Ness M.~K.,  2019, \mn@doi [ApJ] {10.3847/1538-4357/aaf648}, 871, 120

\bibitem[\protect\citeauthoryear{Epanechnikov}{Epanechnikov}{1969}]{epanechnikov_non-parametric_1969}
Epanechnikov V.~A.,  1969, \mn@doi [Theory Probab. Appl.] {10.1137/1114019}, 14, 153

\bibitem[\protect\citeauthoryear{Erb, Steidel, Shapley, Pettini, Reddy  \& Adelberger}{Erb et~al.}{2006}]{erb_stellar_2006}
Erb D.~K.,  Steidel C.~C.,  Shapley A.~E.,  Pettini M.,  Reddy N.~A.,   Adelberger K.~L.,  2006, \mn@doi [ApJ] {10.1086/504891}, 646, 107

\bibitem[\protect\citeauthoryear{Faisst et~al.,}{Faisst et~al.}{2016}]{faisst_rest-uv_2016}
Faisst A.~L.,  et~al., 2016, \mn@doi [ApJ] {10.3847/0004-637X/822/1/29}, 822, 29

\bibitem[\protect\citeauthoryear{{Gaia Collaboration} et~al.,}{{Gaia Collaboration} et~al.}{2016}]{gaia_collaboration_gaia_2016}
{Gaia Collaboration} et~al., 2016, \mn@doi [A\&A] {10.1051/0004-6361/201629512}, 595, A2

\bibitem[\protect\citeauthoryear{{Gaia Collaboration} et~al.,}{{Gaia Collaboration} et~al.}{2018}]{gaia_collaboration_gaia_2018}
{Gaia Collaboration} et~al., 2018, \mn@doi [A\&A] {10.1051/0004-6361/201833051}, 616, A1

\bibitem[\protect\citeauthoryear{{Gaia Collaboration} et~al.,}{{Gaia Collaboration} et~al.}{2023}]{gaia_collaboration_gaia_2023}
{Gaia Collaboration} et~al., 2023, \mn@doi [A\&A] {10.1051/0004-6361/202243940}, 674, A1

\bibitem[\protect\citeauthoryear{Galleti, Federici, Bellazzini, Fusi~Pecci  \& Macrina}{Galleti et~al.}{2004}]{galleti_2mass_2004}
Galleti S.,  Federici L.,  Bellazzini M.,  Fusi~Pecci F.,   Macrina S.,  2004, \mn@doi [A\&A] {10.1051/0004-6361:20035632}, 416, 917

\bibitem[\protect\citeauthoryear{Garrison-Kimmel, Boylan-Kolchin, Bullock  \& Lee}{Garrison-Kimmel et~al.}{2014}]{garrison-kimmel_elvis_2014}
Garrison-Kimmel S.,  Boylan-Kolchin M.,  Bullock J.~S.,   Lee K.,  2014, \mn@doi [MNRAS] {10.1093/mnras/stt2377}, 438, 2578

\bibitem[\protect\citeauthoryear{Genzel et~al.,}{Genzel et~al.}{2015}]{genzel_combined_2015}
Genzel R.,  et~al., 2015, \mn@doi [ApJ] {10.1088/0004-637X/800/1/20}, 800, 20

\bibitem[\protect\citeauthoryear{Gieles \& Baumgardt}{Gieles \& Baumgardt}{2008}]{gieles_lifetimes_2008}
Gieles M.,  Baumgardt H.,  2008, \mn@doi [MNRAS] {10.1111/j.1745-3933.2008.00515.x}, 389, L28

\bibitem[\protect\citeauthoryear{Gieles \& Gnedin}{Gieles \& Gnedin}{2023}]{gieles_mass-loss_2023}
Gieles M.,  Gnedin O.~Y.,  2023, \mn@doi [MNRAS] {10.1093/mnras/stad1287}, 522, 5340

\bibitem[\protect\citeauthoryear{Halbesma, Grand, Gómez, Marinacci, Pakmor, Trick, Busch  \& White}{Halbesma et~al.}{2020}]{halbesma_globular_2020}
Halbesma T. L.~R.,  Grand R. J.~J.,  Gómez F.~A.,  Marinacci F.,  Pakmor R.,  Trick W.~H.,  Busch P.,   White S. D.~M.,  2020, \mn@doi [MNRAS] {10.1093/mnras/staa1380}, 496, 638

\bibitem[\protect\citeauthoryear{Hammer, Puech, Chemin, Flores  \& Lehnert}{Hammer et~al.}{2007}]{hammer_milky_2007}
Hammer F.,  Puech M.,  Chemin L.,  Flores H.,   Lehnert M.~D.,  2007, \mn@doi [ApJ] {10.1086/516727}, 662, 322

\bibitem[\protect\citeauthoryear{Harris}{Harris}{1996}]{harris_catalog_1996}
Harris W.~E.,  1996, \mn@doi [AJ] {10.1086/118116}, 112, 1487

\bibitem[\protect\citeauthoryear{Harris et~al.,}{Harris et~al.}{2020}]{harris_array_2020}
Harris C.~R.,  et~al., 2020, \mn@doi [Nature] {10.1038/s41586-020-2649-2}, 585, 357

\bibitem[\protect\citeauthoryear{Heintz et~al.,}{Heintz et~al.}{2023}]{heintz_dilution_2023}
Heintz K.~E.,  et~al., 2023, arXiv:2212.02890 [astro-ph]

\bibitem[\protect\citeauthoryear{Helmi, Babusiaux, Koppelman, Massari, Veljanoski  \& Brown}{Helmi et~al.}{2018}]{helmi_merger_2018}
Helmi A.,  Babusiaux C.,  Koppelman H.~H.,  Massari D.,  Veljanoski J.,   Brown A. G.~A.,  2018, \mn@doi [Nature] {10.1038/s41586-018-0625-x}, 563, 85

\bibitem[\protect\citeauthoryear{Hilker, Baumgardt, Sollima  \& Bellini}{Hilker et~al.}{2019}]{hilker_galactic_2019}
Hilker M.,  Baumgardt H.,  Sollima A.,   Bellini A.,  2019, \mn@doi [Proc. IAU] {10.1017/S1743921319006823}, 14, 451

\bibitem[\protect\citeauthoryear{Hsiao et~al.,}{Hsiao et~al.}{2023}]{hsiao_jwst_2023}
Hsiao T. Y.-Y.,  et~al., 2023, arXiv:2305.03042 [astro-ph]

\bibitem[\protect\citeauthoryear{Hunter}{Hunter}{2007}]{hunter_matplotlib_2007}
Hunter J.~D.,  2007, \mn@doi [Comput. Sci. Eng.] {10.1109/MCSE.2007.55}, 9, 90

\bibitem[\protect\citeauthoryear{Jung et~al.,}{Jung et~al.}{2023}]{jung_ceers_2023}
Jung I.,  et~al., 2023, arXiv:2304.05385 [astro-ph]

\bibitem[\protect\citeauthoryear{Kewley \& Ellison}{Kewley \& Ellison}{2008}]{kewley_metallicity_2008}
Kewley L.~J.,  Ellison S.~L.,  2008, \mn@doi [ApJ] {10.1086/587500}, 681, 1183

\bibitem[\protect\citeauthoryear{Kravtsov \& Gnedin}{Kravtsov \& Gnedin}{2005}]{kravtsov_formation_2005}
Kravtsov A.~V.,  Gnedin O.~Y.,  2005, \mn@doi [ApJ] {10.1086/428636}, 623, 650

\bibitem[\protect\citeauthoryear{Kravtsov \& Manwadkar}{Kravtsov \& Manwadkar}{2022}]{kravtsov_span_2022}
Kravtsov A.,  Manwadkar V.,  2022, \mn@doi [MNRAS] {10.1093/mnras/stac1439}, 514, 2667

\bibitem[\protect\citeauthoryear{Kravtsov, Klypin  \& Khokhlov}{Kravtsov et~al.}{1997}]{kravtsov_adaptive_1997}
Kravtsov A.~V.,  Klypin A.~A.,   Khokhlov A.~M.,  1997, \mn@doi [ApJS] {10.1086/313015}, 111, 73

\bibitem[\protect\citeauthoryear{Kravtsov, Vikhlinin  \& Meshcheryakov}{Kravtsov et~al.}{2018}]{kravtsov_stellar_2018}
Kravtsov A.~V.,  Vikhlinin A.~A.,   Meshcheryakov A.~V.,  2018, \mn@doi [Astron. Lett.] {10.1134/S1063773717120015}, 44, 8

\bibitem[\protect\citeauthoryear{Kruijssen}{Kruijssen}{2009}]{kruijssen_evolution_2009}
Kruijssen J. M.~D.,  2009, \mn@doi [A\&A] {10.1051/0004-6361/200913325}, 507, 1409

\bibitem[\protect\citeauthoryear{Kruijssen \& Lamers}{Kruijssen \& Lamers}{2008}]{kruijssen_photometric_2008}
Kruijssen J. M.~D.,  Lamers H. J. G. L.~M.,  2008, \mn@doi [A\&A] {10.1051/0004-6361:200810167}, 490, 151

\bibitem[\protect\citeauthoryear{Kruijssen, Pelupessy, Lamers, Portegies~Zwart  \& Icke}{Kruijssen et~al.}{2011}]{kruijssen_modelling_2011}
Kruijssen J. M.~D.,  Pelupessy F.~I.,  Lamers H. J. G. L.~M.,  Portegies~Zwart S.~F.,   Icke V.,  2011, \mn@doi [MNRAS] {10.1111/j.1365-2966.2011.18467.x}, 414, 1339

\bibitem[\protect\citeauthoryear{Kruijssen, Pfeffer, Crain  \& Bastian}{Kruijssen et~al.}{2019}]{kruijssen_e-mosaics_2019}
Kruijssen J. M.~D.,  Pfeffer J.~L.,  Crain R.~A.,   Bastian N.,  2019, \mn@doi [MNRAS] {10.1093/mnras/stz968}, 486, 3134

\bibitem[\protect\citeauthoryear{Leitner}{Leitner}{2012}]{leitner_last_2012}
Leitner S.~N.,  2012, \mn@doi [ApJ] {10.1088/0004-637X/745/2/149}, 745, 149

\bibitem[\protect\citeauthoryear{Lewis et~al.,}{Lewis et~al.}{2023}]{lewis_gas-phase_2023}
Lewis Z.~J.,  et~al., 2023, arXiv:2304.12343 [astro-ph]

\bibitem[\protect\citeauthoryear{Li \& Gnedin}{Li \& Gnedin}{2014}]{li_modeling_2014}
Li H.,  Gnedin O.~Y.,  2014, \mn@doi [ApJ] {10.1088/0004-637X/796/1/10}, 796, 10

\bibitem[\protect\citeauthoryear{Li \& Gnedin}{Li \& Gnedin}{2019}]{li_star_2019}
Li H.,  Gnedin O.~Y.,  2019, \mn@doi [MNRAS] {10.1093/mnras/stz1114}, 486, 4030

\bibitem[\protect\citeauthoryear{Li, Gnedin, Gnedin, Meng, Semenov  \& Kravtsov}{Li et~al.}{2017}]{li_star_2017}
Li H.,  Gnedin O.~Y.,  Gnedin N.~Y.,  Meng X.,  Semenov V.~A.,   Kravtsov A.~V.,  2017, \mn@doi [ApJ] {10.3847/1538-4357/834/1/69}, 834, 69

\bibitem[\protect\citeauthoryear{Li, Gnedin  \& Gnedin}{Li et~al.}{2018}]{li_star_2018}
Li H.,  Gnedin O.~Y.,   Gnedin N.~Y.,  2018, \mn@doi [ApJ] {10.3847/1538-4357/aac9b8}, 861, 107

\bibitem[\protect\citeauthoryear{Li et~al.,}{Li et~al.}{2022}]{li_mass-metallicity_2022}
Li M.,  et~al., 2022, arXiv:2211.01382 [astro-ph]

\bibitem[\protect\citeauthoryear{Lilly, Carollo, Pipino, Renzini  \& Peng}{Lilly et~al.}{2013}]{lilly_gas_2013}
Lilly S.~J.,  Carollo C.~M.,  Pipino A.,  Renzini A.,   Peng Y.,  2013, \mn@doi [ApJ] {10.1088/0004-637X/772/2/119}, 772, 119

\bibitem[\protect\citeauthoryear{Mackey et~al.,}{Mackey et~al.}{2019}]{mackey_two_2019}
Mackey D.,  et~al., 2019, \mn@doi [Nature] {10.1038/s41586-019-1597-1}, 574, 69

\bibitem[\protect\citeauthoryear{Maiolino et~al.,}{Maiolino et~al.}{2008}]{maiolino_amaze_2008}
Maiolino R.,  et~al., 2008, \mn@doi [A\&A] {10.1051/0004-6361:200809678}, 488, 463

\bibitem[\protect\citeauthoryear{Majewski et~al.,}{Majewski et~al.}{2017}]{majewski_apache_2017}
Majewski S.~R.,  et~al., 2017, \mn@doi [AJ] {10.3847/1538-3881/aa784d}, 154, 94

\bibitem[\protect\citeauthoryear{Malhan et~al.,}{Malhan et~al.}{2022}]{malhan_global_2022}
Malhan K.,  et~al., 2022, \mn@doi [ApJ] {10.3847/1538-4357/ac4d2a}, 926, 107

\bibitem[\protect\citeauthoryear{Mannucci et~al.,}{Mannucci et~al.}{2009}]{mannucci_lsd_2009}
Mannucci F.,  et~al., 2009, \mn@doi [MNRAS] {10.1111/j.1365-2966.2009.15185.x}, 398, 1915

\bibitem[\protect\citeauthoryear{Massari, Koppelman  \& Helmi}{Massari et~al.}{2019}]{massari_origin_2019}
Massari D.,  Koppelman H.~H.,   Helmi A.,  2019, \mn@doi [A\&A] {10.1051/0004-6361/201936135}, 630, L4

\bibitem[\protect\citeauthoryear{Matthee, Mackenzie, Simcoe, Kashino, Lilly, Bordoloi  \& Eilers}{Matthee et~al.}{2023}]{matthee_eiger_2023}
Matthee J.,  Mackenzie R.,  Simcoe R.~A.,  Kashino D.,  Lilly S.~J.,  Bordoloi R.,   Eilers A.-C.,  2023, \mn@doi [ApJ] {10.3847/1538-4357/acc846}, 950, 67

\bibitem[\protect\citeauthoryear{McMillan}{McMillan}{2017}]{mcmillan_mass_2017}
McMillan P.~J.,  2017, \mn@doi [MNRAS] {10.1093/mnras/stw2759}, 465, 76

\bibitem[\protect\citeauthoryear{Muratov \& Gnedin}{Muratov \& Gnedin}{2010}]{muratov_modeling_2010}
Muratov A.~L.,  Gnedin O.~Y.,  2010, \mn@doi [ApJ] {10.1088/0004-637X/718/2/1266}, 718, 1266

\bibitem[\protect\citeauthoryear{Mészáros et~al.,}{Mészáros et~al.}{2020}]{meszaros_homogeneous_2020}
Mészáros S.,  et~al., 2020, \mn@doi [MNRAS] {10.1093/mnras/stz3496}, 492, 1641

\bibitem[\protect\citeauthoryear{Mészáros et~al.,}{Mészáros et~al.}{2021}]{meszaros_homogeneous_2021}
Mészáros S.,  et~al., 2021, \mn@doi [MNRAS] {10.1093/mnras/stab1208}, 505, 1645

\bibitem[\protect\citeauthoryear{Nakajima, Ouchi, Isobe, Harikane, Zhang, Ono, Umeda  \& Oguri}{Nakajima et~al.}{2023}]{nakajima_jwst_2023}
Nakajima K.,  Ouchi M.,  Isobe Y.,  Harikane Y.,  Zhang Y.,  Ono Y.,  Umeda H.,   Oguri M.,  2023, arXiv:2301.12825 [astro-ph]

\bibitem[\protect\citeauthoryear{Nelson et~al.,}{Nelson et~al.}{2019}]{nelson_first_2019}
Nelson D.,  et~al., 2019, \mn@doi [MNRAS] {10.1093/mnras/stz2306}, 490, 3234

\bibitem[\protect\citeauthoryear{Nelson et~al.,}{Nelson et~al.}{2021}]{nelson_illustristng_2021}
Nelson D.,  et~al., 2021, arXiv:1812.05609 [astro-ph]

\bibitem[\protect\citeauthoryear{Pancino et~al.,}{Pancino et~al.}{2017}]{pancino_gaia_2017}
Pancino E.,  et~al., 2017, \mn@doi [A\&A] {10.1051/0004-6361/201730474}, 601, A112

\bibitem[\protect\citeauthoryear{Peng et~al.,}{Peng et~al.}{2006}]{peng_acs_2006}
Peng E.~W.,  et~al., 2006, \mn@doi [ApJ] {10.1086/498210}, 639, 95

\bibitem[\protect\citeauthoryear{Pfeffer, Kruijssen, Crain  \& Bastian}{Pfeffer et~al.}{2018}]{pfeffer_e-mosaics_2018}
Pfeffer J.,  Kruijssen J. M.~D.,  Crain R.~A.,   Bastian N.,  2018, \mn@doi [MNRAS] {10.1093/mnras/stx3124}, 475, 4309

\bibitem[\protect\citeauthoryear{Pfeffer, Trujillo-Gomez, Kruijssen, Crain, Hughes, Reina-Campos  \& Bastian}{Pfeffer et~al.}{2020}]{pfeffer_predicting_2020}
Pfeffer J.~L.,  Trujillo-Gomez S.,  Kruijssen J. M.~D.,  Crain R.~A.,  Hughes M.~E.,  Reina-Campos M.,   Bastian N.,  2020, \mn@doi [MNRAS] {10.1093/mnras/staa3109}, 499, 4863

\bibitem[\protect\citeauthoryear{Phipps, Khochfar, Lisa~Varri  \& Dalla~Vecchia}{Phipps et~al.}{2020}]{phipps_first_2020}
Phipps F.,  Khochfar S.,  Lisa~Varri A.,   Dalla~Vecchia C.,  2020, \mn@doi [A\&A] {10.1051/0004-6361/202037884}, 641, A132

\bibitem[\protect\citeauthoryear{Pillepich et~al.,}{Pillepich et~al.}{2018}]{pillepich_simulating_2018}
Pillepich A.,  et~al., 2018, \mn@doi [MNRAS] {10.1093/mnras/stx2656}, 473, 4077

\bibitem[\protect\citeauthoryear{Pillepich et~al.,}{Pillepich et~al.}{2019}]{pillepich_first_2019}
Pillepich A.,  et~al., 2019, \mn@doi [MNRAS] {10.1093/mnras/stz2338}, 490, 3196

\bibitem[\protect\citeauthoryear{Pillepich et~al.,}{Pillepich et~al.}{2023}]{pillepich_milky_2023}
Pillepich A.,  et~al., 2023, arXiv:2303.16217 [astro-ph]

\bibitem[\protect\citeauthoryear{{Planck Collaboration} et~al.,}{{Planck Collaboration} et~al.}{2016}]{planck_collaboration_planck_2016}
{Planck Collaboration} et~al., 2016, \mn@doi [A\&A] {10.1051/0004-6361/201525830}, 594, A13

\bibitem[\protect\citeauthoryear{Reina-Campos, Hughes, Kruijssen, Pfeffer, Bastian, Crain, Koch  \& Grebel}{Reina-Campos et~al.}{2020}]{reina-campos_mass_2020}
Reina-Campos M.,  Hughes M.~E.,  Kruijssen J. M.~D.,  Pfeffer J.~L.,  Bastian N.,  Crain R.~A.,  Koch A.,   Grebel E.~K.,  2020, \mn@doi [MNRAS] {10.1093/mnras/staa483}, 493, 3422

\bibitem[\protect\citeauthoryear{Reina-Campos, Keller, Kruijssen, Gensior, Trujillo-Gomez, Jeffreson, Pfeffer  \& Sills}{Reina-Campos et~al.}{2022}]{reina-campos_introducing_2022}
Reina-Campos M.,  Keller B.~W.,  Kruijssen J. M.~D.,  Gensior J.,  Trujillo-Gomez S.,  Jeffreson S. M.~R.,  Pfeffer J.~L.,   Sills A.,  2022, \mn@doi [MNRAS] {10.1093/mnras/stac1934}, 517, 3144

\bibitem[\protect\citeauthoryear{Renaud, Gieles  \& Boily}{Renaud et~al.}{2011}]{renaud_evolution_2011}
Renaud F.,  Gieles M.,   Boily C.~M.,  2011, \mn@doi [MNRAS] {10.1111/j.1365-2966.2011.19531.x}, 418, 759

\bibitem[\protect\citeauthoryear{Renaud, Agertz  \& Gieles}{Renaud et~al.}{2017}]{renaud_origin_2017}
Renaud F.,  Agertz O.,   Gieles M.,  2017, \mn@doi [MNRAS] {10.1093/mnras/stw2969}, 465, 3622

\bibitem[\protect\citeauthoryear{Rodriguez-Gomez et~al.,}{Rodriguez-Gomez et~al.}{2015}]{rodriguez-gomez_merger_2015}
Rodriguez-Gomez V.,  et~al., 2015, \mn@doi [MNRAS] {10.1093/mnras/stv264}, 449, 49

\bibitem[\protect\citeauthoryear{Sanders et~al.,}{Sanders et~al.}{2021}]{sanders_mosdef_2021}
Sanders R.~L.,  et~al., 2021, \mn@doi [ApJ] {10.3847/1538-4357/abf4c1}, 914, 19

\bibitem[\protect\citeauthoryear{Sanders et~al.,}{Sanders et~al.}{2023}]{sanders_co_2023}
Sanders R.~L.,  et~al., 2023, \mn@doi [ApJ] {10.3847/1538-4357/aca46f}, 942, 24

\bibitem[\protect\citeauthoryear{Savaglio et~al.,}{Savaglio et~al.}{2005}]{savaglio_gemini_2005}
Savaglio S.,  et~al., 2005, \mn@doi [ApJ] {10.1086/497331}, 635, 260

\bibitem[\protect\citeauthoryear{Schaye et~al.,}{Schaye et~al.}{2015}]{schaye_eagle_2015}
Schaye J.,  et~al., 2015, \mn@doi [MNRAS] {10.1093/mnras/stu2058}, 446, 521

\bibitem[\protect\citeauthoryear{Schechter}{Schechter}{1976}]{schechter_analytic_1976}
Schechter P.,  1976, \mn@doi [ApJ] {10.1086/154079}, 203, 297

\bibitem[\protect\citeauthoryear{Semenov, Conroy, Chandra, Hernquist  \& Nelson}{Semenov et~al.}{2023}]{semenov_formation_2023}
Semenov V.~A.,  Conroy C.,  Chandra V.,  Hernquist L.,   Nelson D.,  2023, arXiv: 2306.09398 [astro-ph]

\bibitem[\protect\citeauthoryear{Simien, Athanassoula, Pellet, Monnet, Maucherat  \& Courtès}{Simien et~al.}{1978}]{simien_spiral_1978}
Simien F.,  Athanassoula E.,  Pellet A.,  Monnet G.,  Maucherat A.,   Courtès G.,  1978, A\&A, 67, 73

\bibitem[\protect\citeauthoryear{Springel}{Springel}{2010}]{springel_e_2010}
Springel V.,  2010, \mn@doi [MNRAS] {10.1111/j.1365-2966.2009.15715.x}, 401, 791

\bibitem[\protect\citeauthoryear{Springel, White, Tormen  \& Kauffmann}{Springel et~al.}{2001}]{springel_populating_2001}
Springel V.,  White S. D.~M.,  Tormen G.,   Kauffmann G.,  2001, \mn@doi [MNRAS] {10.1046/j.1365-8711.2001.04912.x}, 328, 726

\bibitem[\protect\citeauthoryear{Strom, Rudie, Steidel  \& Trainor}{Strom et~al.}{2022}]{strom_chemical_2022}
Strom A.~L.,  Rudie G.~C.,  Steidel C.~C.,   Trainor R.~F.,  2022, \mn@doi [ApJ] {10.3847/1538-4357/ac38a3}, 925, 116

\bibitem[\protect\citeauthoryear{Tacconi et~al.,}{Tacconi et~al.}{2018}]{tacconi_phibss_2018}
Tacconi L.~J.,  et~al., 2018, \mn@doi [ApJ] {10.3847/1538-4357/aaa4b4}, 853, 179

\bibitem[\protect\citeauthoryear{Topping et~al.,}{Topping et~al.}{2021}]{topping_mosdef_2021}
Topping M.~W.,  et~al., 2021, \mn@doi [MNRAS] {10.1093/mnras/stab1793}, 506, 1237

\bibitem[\protect\citeauthoryear{Tremonti et~al.,}{Tremonti et~al.}{2004}]{tremonti_origin_2004}
Tremonti C.~A.,  et~al., 2004, \mn@doi [ApJ] {10.1086/423264}, 613, 898

\bibitem[\protect\citeauthoryear{Trujillo-Gomez, Diederik Kruijssen, Reina-Campos, Pfeffer, Keller, Crain, Bastian  \& Hughes}{Trujillo-Gomez et~al.}{2021}]{trujillo-gomez_kinematics_2021}
Trujillo-Gomez S.,  Diederik Kruijssen J.~M.,  Reina-Campos M.,  Pfeffer J.~L.,  Keller B.~W.,  Crain R.~A.,  Bastian N.,   Hughes M.~E.,  2021, \mn@doi [MNRAS] {10.1093/mnras/stab341}, 503, 31

\bibitem[\protect\citeauthoryear{Turk, Smith, Oishi, Skory, Skillman, Abel  \& Norman}{Turk et~al.}{2011}]{turk_yt_2011}
Turk M.~J.,  Smith B.~D.,  Oishi J.~S.,  Skory S.,  Skillman S.~W.,  Abel T.,   Norman M.~L.,  2011, \mn@doi [ApJS] {10.1088/0067-0049/192/1/9}, 192, 9

\bibitem[\protect\citeauthoryear{Valenzuela, Moster, Remus, O’Leary  \& Burkert}{Valenzuela et~al.}{2021}]{valenzuela_globular_2021}
Valenzuela L.~M.,  Moster B.~P.,  Remus R.-S.,  O’Leary J.~A.,   Burkert A.,  2021, \mn@doi [MNRAS] {10.1093/mnras/stab1701}, 505, 5815

\bibitem[\protect\citeauthoryear{Valenzuela, Remus, McKenzie  \& Forbes}{Valenzuela et~al.}{2023}]{valenzuela_galaxy_2023}
Valenzuela L.~M.,  Remus R.-S.,  McKenzie M.,   Forbes D.~A.,  2023, arXiv:2309.11545 [astro-ph]

\bibitem[\protect\citeauthoryear{Vasiliev}{Vasiliev}{2019}]{vasiliev_agama_2019}
Vasiliev E.,  2019, \mn@doi [MNRAS] {10.1093/mnras/sty2672}, 482, 1525

\bibitem[\protect\citeauthoryear{Vasiliev, Belokurov  \& Erkal}{Vasiliev et~al.}{2021}]{vasiliev_tango_2021}
Vasiliev E.,  Belokurov V.,   Erkal D.,  2021, \mn@doi [MNRAS] {10.1093/mnras/staa3673}, 501, 2279

\bibitem[\protect\citeauthoryear{Vazdekis, Sánchez-Blázquez, Falcón-Barroso, Cenarro, Beasley, Cardiel, Gorgas  \& Peletier}{Vazdekis et~al.}{2010}]{vazdekis_evolutionary_2010}
Vazdekis A.,  Sánchez-Blázquez P.,  Falcón-Barroso J.,  Cenarro A.~J.,  Beasley M.~A.,  Cardiel N.,  Gorgas J.,   Peletier R.~F.,  2010, \mn@doi [MNRAS] {10.1111/j.1365-2966.2010.16407.x}

\bibitem[\protect\citeauthoryear{Virtanen et~al.,}{Virtanen et~al.}{2020}]{virtanen_scipy_2020}
Virtanen P.,  et~al., 2020, \mn@doi [Nat Methods] {10.1038/s41592-019-0686-2}, 17, 261

\bibitem[\protect\citeauthoryear{Wang et~al.,}{Wang et~al.}{2022}]{wang_3_2022}
Wang T.-M.,  et~al., 2022, \mn@doi [A\&A] {10.1051/0004-6361/202142299}, 660, A142

\bibitem[\protect\citeauthoryear{Williams et~al.,}{Williams et~al.}{2023}]{williams_magnified_2023}
Williams H.,  et~al., 2023, \mn@doi [Science] {10.1126/science.adf5307}, 380, 416

\bibitem[\protect\citeauthoryear{Yu et~al.,}{Yu et~al.}{2023}]{yu_born_2023}
Yu S.,  et~al., 2023, \mn@doi [MNRAS] {10.1093/mnras/stad1806}, 523, 6220

\bibitem[\protect\citeauthoryear{Zahid, Kewley  \& Bresolin}{Zahid et~al.}{2011}]{zahid_mass-metallicity_2011}
Zahid H.~J.,  Kewley L.~J.,   Bresolin F.,  2011, \mn@doi [ApJ] {10.1088/0004-637X/730/2/137}, 730, 137

\bibitem[\protect\citeauthoryear{Zahid, Dima, Kudritzki, Kewley, Geller, Hwang, Silverman  \& Kashino}{Zahid et~al.}{2014}]{zahid_universal_2014}
Zahid H.~J.,  Dima G.~I.,  Kudritzki R.-P.,  Kewley L.~J.,  Geller M.~J.,  Hwang H.~S.,  Silverman J.~D.,   Kashino D.,  2014, \mn@doi [ApJ] {10.1088/0004-637X/791/2/130}, 791, 130

\makeatother
\end{thebibliography}

%%%%%%%%%%%%%%%%% APPENDICES %%%%%%%%%%%%%%%%%%%%%

\appendix

\section{The redshift-dependent mass--metallicity relation}
\label{sec:mzr}

We calculate the metallicity of GCs using a redshift-dependent MZR. Before introducing the MZR in depth, we first introduce the definition of metallicity. Traditionally, metallicity is expressed in the logarithmic abundance ratio $\log(\rm{X/Y})$, which is short for $\log(N_{\rm{X}}/N_{\rm{Y}})$. A common practice is to normalize this value with a fixed offset of 12 (e.g., $\log\epsilon_{\rm X}\equiv12+\log(\rm{X/H})$), or with the solar value: ${\rm [X/Y]}\equiv\log(\rm{X/Y}) - \log(\rm{X/Y})_\odot$. The former is often used to describe gas-phase metallicity, while the latter is more frequently used for stellar metallicity. In this work, we use the solar reference values from \citet{asplund_chemical_2021}:
\begin{equation}
    \begin{split}
        12+\log(\rm{O/H})_\odot &= 8.69 \\
        12+\log(\rm{Fe/H})_\odot &= 7.46.
    \end{split}
    \label{eq:solar}
\end{equation}

The MZR used to be poorly constrained for dwarf galaxies ($\Mstar\lesssim10^8\Msun$) and in the early universe ($z\gtrsim 6$). The launch of JWST has significantly extended the observed galaxy metallicity in both directions, enabling us to re-calibrate the MZR. We use the following factorized power-law relation:
\begin{equation}
    \feh = \alpha_M\log\frac{\Mstar}{M_0} - \alpha_z\log(1+z) + \feh_0
    \label{eq:mzr}
\end{equation}
where $\alpha_M$ and $\alpha_z$ are the logarithmic slopes of the stellar mass and redshift dependence, respectively. $\feh_0$ is the metallicity at some characteristic mass scale $M_0$ at $z=0$. We take $M_0=10^9\Msun$ to represent the typical mass of host galaxies at the formation time of the majority of surviving clusters. Note that such a simple power-law scaling is not valid for most massive galaxies ($\Mstar\gtrsim10^{11}\Msun$), where metallicity saturates at a slight supersolar value \citep[see, e.g.,][]{tremonti_origin_2004}. Therefore, we follow \citet{choksi_formation_2018} to cap the maximum value $\feh_{\rm max} = +0.3$~dex to avoid unreasonably large $\feh$ of GCs in massive galaxies.

\begin{figure}
    \includegraphics[width=\linewidth]{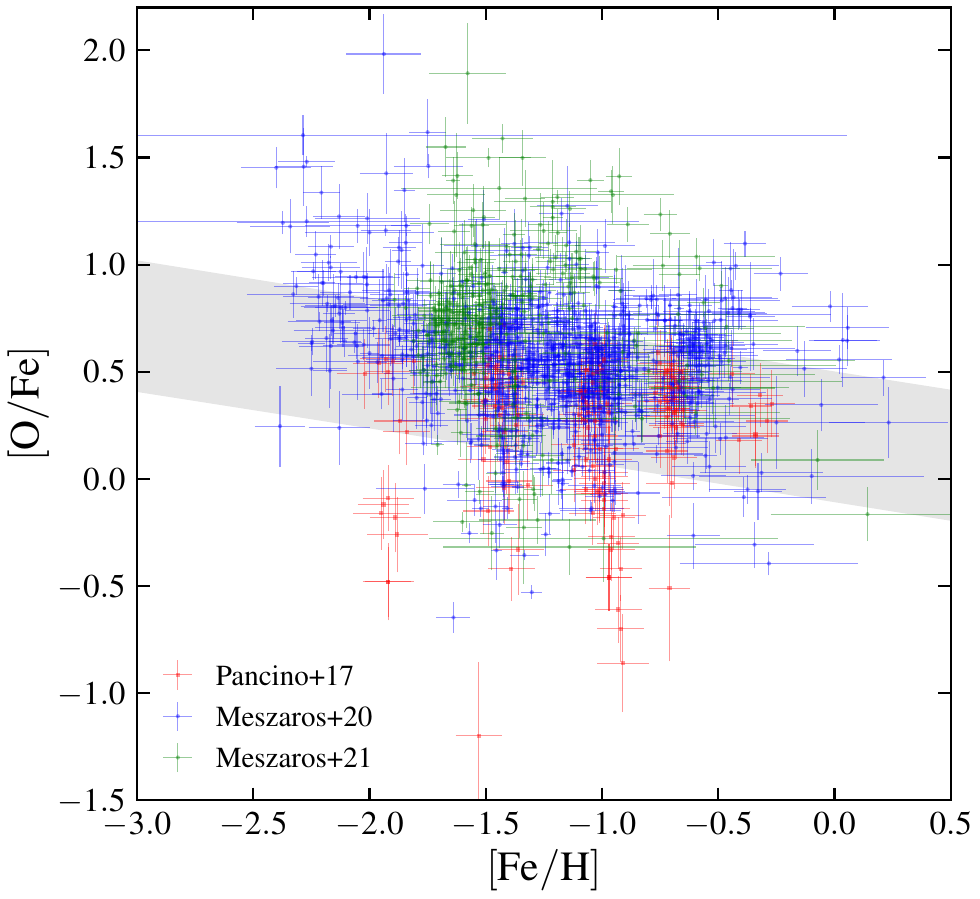}
    \vspace{-4mm}
    \caption{$\ofe$--$\feh$ relation of stars in nearby GCs. We plot the GC stars from the Gaia-ESO survey as red symbols \citep{pancino_gaia_2017} and the SDSS-IV APOGEE-2 survey as blue \citep[excluding $\omega$ Cen,][]{meszaros_homogeneous_2020} and green symbols \citep[$\omega$ Cen only,][]{meszaros_homogeneous_2021} with errorbars representing the measurement uncertainty. We show the linear fit of the combined data as the gray shaded region with a 0.3-dex scatter.}
    \label{fig:ofe_feh}
\end{figure}

We assume that the metallicity of GCs is directly inherited from the surrounding gas at formation. Since most measurements of the gas-phase metallicity in high-redshift galaxies provide the oxygen abundance rather than the iron abundance, we need to convert the oxygen abundance to $\feh$ using the observed $\ofe$--$\feh$ relation. Combining the observational data of nearby GCs from the Gaia-ESO survey \citep{pancino_gaia_2017} and the SDSS-IV APOGEE-2 survey \citep{meszaros_homogeneous_2020,meszaros_homogeneous_2021}, we find an anti-correlation between $\ofe$ and $\feh$, see Fig.~\ref{fig:ofe_feh}. This relation can be quantified by a linear relation: $\ofe = 0.37-0.17\feh$, with a 0.3-dex scatter, in the range of $-2\lesssim\feh\lesssim0$. We emphasize that this relation is only valid for stars in GCs as these works only analysed stellar abundances in nearby GCs. 

Given $\log({\rm O/Fe})$, we can now convert the measured oxygen abundance to the gas-phase iron abundance:
\begin{equation}
    \log({\rm Fe/H}) = \log({\rm O/H}) - \log({\rm O/Fe}).
\end{equation}
We then rewrite this equation in terms of $\feh$ and $\ofe$:
\begin{equation}
    \feh = \log({\rm O/H}) - \ofe - \log(\rm{O/H})_\odot.
    \label{eq:ofe2feh}
\end{equation}
Combining Eqs.~(\ref{eq:solar}), (\ref{eq:ofe2feh}), and the linear anti-correlation between $\ofe$ and $\feh$, we obtain
\begin{equation}
    \feh = 1.20\, (12+\log({\rm O/H})) - 10.92.
    \label{eq:o2feh}
\end{equation}

Although the MZR is well constrained for nearby ($z\sim0$) galaxies across a wide range of stellar mass, we must calibrate the parameters of Eq.~(\ref{eq:mzr}) specifically at earlier times when GC formation is most active. Therefore, we incorporate observational gas-phase MZR from $z\sim0.7$ up to $z\sim12 $ from multiple surveys as described below.

\citet{maiolino_amaze_2008} combined MZR measurements from earlier works \citep{tremonti_origin_2004,kewley_metallicity_2008,savaglio_gemini_2005,erb_stellar_2006} at $z\sim0-3$ and the Assessing the Mass-Abundance redshift[-Z] Evolution (AMAZE) program at $z\sim3.5$. These authors used several emission line ratios as metallicity diagnostics to derive the gas-phase metallicity. Note that their data at $z\sim3.5$ are primarily star forming Lyman-break galaxies, which may introduce a systematic bias in the derived the MZR. 

\citet{zahid_universal_2014} put together observational data from the Sloan Digital Sky Survey (SDSS), Smithsonian Hectospec Lensing Survey (SHELS), Deep Extragalactic Evolutionary Probe 2 (DEEP2), and the Fiber Multi-Object Spectrograph-Cosmic Evolution Survey (FMOS-COSMOS) spanning a redshift range $0<z<1.7$. They obtained the gas-phase oxygen abundance via metallicity--line ratio scaling.

\citet{lewis_gas-phase_2023} measured oxygen abundances for 145 galaxies at $z\sim0.7$. They estimated the metallicity via a Bayesian framework using emission lines from the Large Early Galaxy Astrophysics Census (LEGA-C) survey.

\citet{strom_chemical_2022} calculated the gas-phase metallicity of a subsample of the Keck Baryonic Structure Survey (KBSS) galaxies at $z\sim2-3$. They tested various methods to determine the oxygen abundance and found noticeable discrepancy. Here, we apply the metallicity indicated by their photoionization model.

\citet{li_mass-metallicity_2022} measured the metallicity of 55 galaxies in the Abell 2744 (by the GLASS JWST Early Release Science program) and SMACS J0723-3732 (by the JWST Early Release Observations program) galaxy cluster fields. These galaxies span a redshift range $z=2-3$. They also employed metallicity--line ratio scaling to derive the gas phase metallicity.

\citet{sanders_mosdef_2021} analysed the metallicity for about 300 galaxies at $z\sim 2.3$ and 150 galaxies at $z\sim 3.3$ in the MOSFIRE Deep Evolution Field (MOSDEF) survey. They obtained the oxygen abundance first by fitting various line ratios as a function of $\log({\rm O/H})$, and then searching for the best-fit oxygen abundance using the $\chi^2$ minimization technique. 

\citet{sanders_co_2023} incorporated the earlier data \citep{curti_massmetallicity_2020,zahid_mass-metallicity_2011,topping_mosdef_2021,sanders_mosdef_2021} and fitted a functional form to match the MZR from $z=0.08$ to $3.3$. 

\citet{curti_jades_2023} calculated the gas-phase metallicity of 66 galaxies at $z=3-10$. These galaxies are observed with the JWST/NIRSpec as part of the JWST Advanced Deep Extragalactic Survey (JADES). These authors also used line ratio fitting to obtain metallicity via the minimal likelihood technique. They presented their results in combination with \citet{nakajima_jwst_2023}.

\citet{nakajima_jwst_2023} focused on 135 galaxies at $z=4-10$ from the JWST/NIRSpec data by the ERO, GLASS, and Cosmic Evolution Early Release Science (CEERS) programs. They used a combination of direct $T_e$ (electron temperature) method and metallicity--line ratio scaling to estimate $\log({\rm O/H})$.

\citet{faisst_rest-uv_2016} measured the metallicity of 224 galaxies at $z\sim5$ from COSMOS. They estimated the gas-phase oxygen abundance using the ultraviolet equivalent width--metallicity correlation. Their sample includes Ly$\alpha$ emitting galaxies and galaxies without Ly$\alpha$ emission. Here, we only consider the average MZR of both categories.

\citet{matthee_eiger_2023} analysed 117 galaxies at $z=5.33-6.93$ from the Emission-line galaxies and the Intergalactic Gasinthe Epoch of Reionization (EIGER) program using JWST/NIRSpec. These authors employed photo-ionization modeling for spectral energy distribution (SED) fitting to calculate the gas-phase metallicity along with the stellar mass and other model parameters.

\citet{heintz_dilution_2023} measured the metallicity of 16 galaxies at $z=7-10$ with JWST/NIRSpec from the gravitational lensing clusters Abell 2744 and RXJ-2129, and the CEERS survey. They also determined the metallicity using direct $T_e$ method and metallicity--line ratio scaling.

In addition to the above surveys, we also include JWST measurements from \citet[][two galaxies at $z\sim4$]{birkin_jwsts_2023}, \citet[][three galaxies at $z=7.47-7.75$]{jung_ceers_2023}, \citet[][three galaxies at $z\sim 8$]{curti_chemical_2022}, \citet[][a galaxy at $z=9.31$]{boyett_massive_2023}, \citet[][a galaxy at $z=9.51$]{williams_magnified_2023}, and \citet[][a galaxy at $z=11.7$]{hsiao_jwst_2023}. 

\begin{figure*}
    \includegraphics[width=0.9\linewidth]{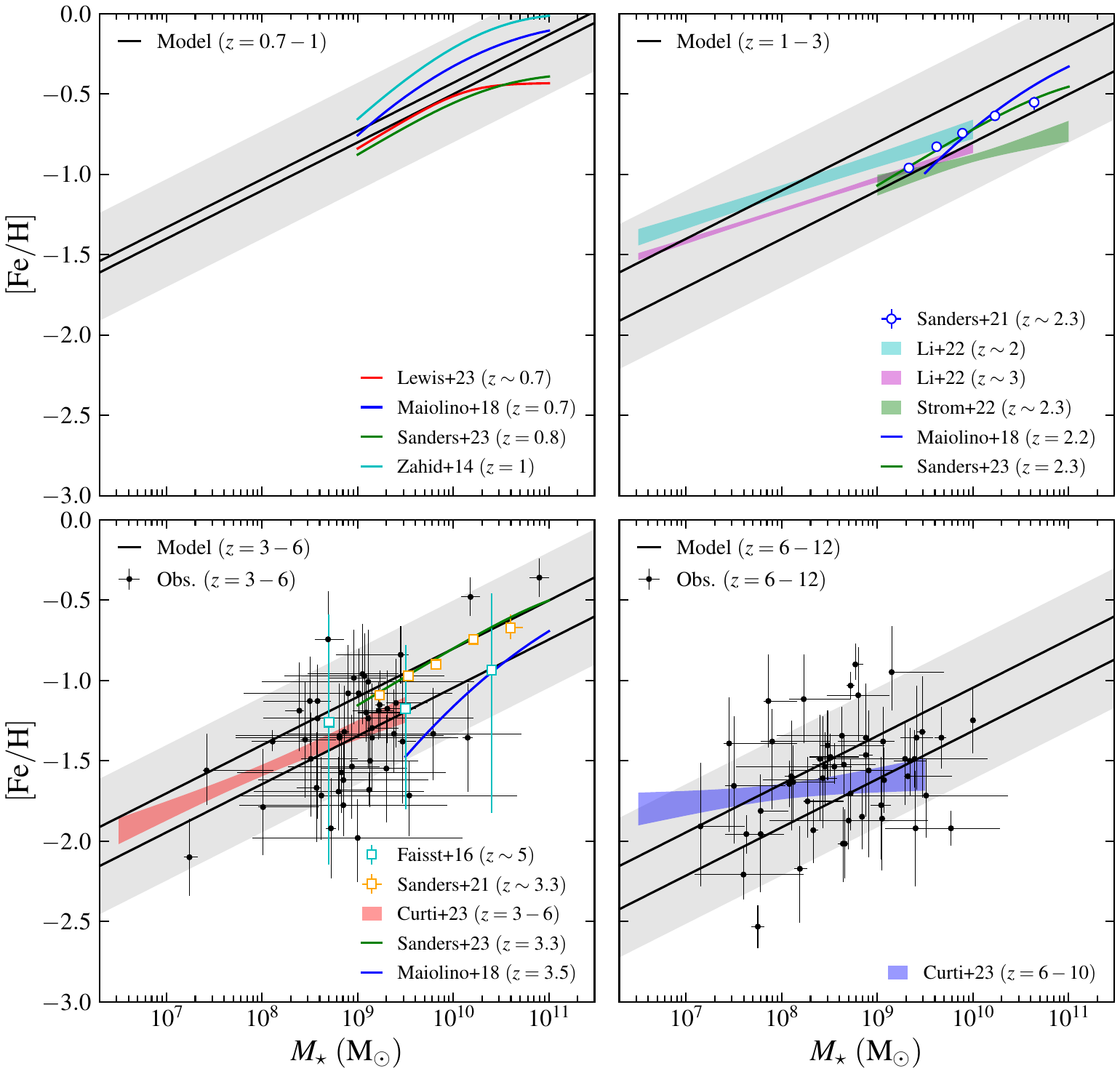}
    \vspace{-1mm}
    \caption{Galaxy stellar mass--gas-phase metallicity relations in different redshift ranges: $z=0.7-1$ (\textit{upper left}), $1-3$ (\textit{upper right}), $3-6$ (\textit{lower left}), and $6-12$ (\textit{lower right}). Points with black errorbars show data for individual galaxies \citep{jung_ceers_2023,curti_chemical_2022,boyett_massive_2023,williams_magnified_2023,hsiao_jwst_2023}. The observational compilations that fit the data with functional forms are shown as colored curves or bands \citep{maiolino_amaze_2008,zahid_universal_2014,lewis_gas-phase_2023,strom_chemical_2022,li_mass-metallicity_2022,sanders_mosdef_2021,curti_jades_2023,nakajima_jwst_2023,faisst_rest-uv_2016,matthee_eiger_2023,heintz_dilution_2023}. We convert the observed oxygen abundance $12+\log({\rm O/H})$ to iron abundance $\feh$ using Eq.~(\ref{eq:o2feh}). The mean MZR adopted in this paper is represented by two black lines in each panel, evaluated at the lower/upper redshift bounds of the panel. The gray shaded ranges show the scatter $\sigma_{\rm g}=0.3$~dex around the bounding mean relations.}
    \label{fig:MZR}
\end{figure*}

In Fig.~\ref{fig:MZR}, we plot these observational data grouped by redshift bins. Galaxies at different redshifts and from different surveys show slopes of the $\feh$--$\Mstar$ relation between 0.2 and 0.4. We find a single value $\sim 0.3$ can match the observational data at $1\lesssim z\lesssim6$ when our model forms majority of surviving clusters. We therefore set $\alpha_M=0.3$ for the model MZR.

In addition, we find a dependence on redshift from $z\sim0.7$ to $z\sim12$, which can be characterized by $\alpha_z=1.0$. This value is slightly greater than what was employed in the previous version of the model, $\alpha_z=0.9$, which accounts for the approximately 0.6~dex drop of metallicity from $z=0$ to $z=4$ at a fixed mass \citep{mannucci_lsd_2009}. 

Finally, we normalize the model MZR by selecting $\feh_0$ that brings Eq.~(\ref{eq:mzr}) to the correct scale. The resulting functional form for the shape and evolution of MZR is
\begin{equation}
    \feh = 0.3\log\frac{\Mstar}{10^9\Msun} - 1.0\log(1+z) - 0.5.
\end{equation}

We plot the final relation in Fig.~\ref{fig:MZR} for comparison with the data. Our MZR can match the observations across a broad mass range ($\Mstar=10^7-10^{11}\Msun$) and redshift range ($z=0.7-12$). The standard deviation of the difference between the observed metallicity of individual galaxies and the predicted metallicity by our MZR is around 0.3~dex. We apply this value as the scatter of mean galaxy metallicity, $\sigma_{\rm g}=0.3$~dex.

% Don't change these lines
\bsp	% typesetting comment
\label{lastpage}
\end{document}